%% file: ANA-SUSY-2020-28-PAPER.tex
\documentclass[cernpreprint, atlasdraft=false, coverpage=false, texlive=2023, USenglish, orcidlogo, texmf, orcidlogo]{atlasdoc}

\usepackage{atlaspackage}
\usepackage{atlasbiblatex}
 
\usepackage{atlasphysics}

\graphicspath{{logos/}{figures/}}
 
\usepackage{ANA-SUSY-2020-28-PAPER-defs}

\AtlasTitle{Search in diphoton and dielectron final states
for displaced production of Higgs or $Z$ bosons
with the ATLAS detector in $\sqrt{s} =
13$ TeV $pp$ collisions }
 
\AtlasVersion{2.6}
 
\AtlasAbstract{

A search is presented for displaced production of Higgs
bosons or $Z$ bosons, originating from the decay of a neutral long-lived particle (LLP) and reconstructed in the decay modes $H\rightarrow \gamma\gamma$ and $Z\rightarrow ee$.
The analysis uses the full Run 2 data set of proton--proton collisions delivered by the LHC at an energy of $\sqrt{s}=13$ TeV between 2015 and 2018 and recorded by the ATLAS detector, corresponding to an integrated luminosity of 139 fb$^{-1}$.
Exploiting the capabilities of the ATLAS liquid argon calorimeter to precisely measure the arrival times and trajectories of electromagnetic objects, the analysis searches for the signature of pairs of photons or electrons which arise from a common displaced vertex and which arrive after some delay at the calorimeter.
The results are interpreted in a gauge-mediated supersymmetry breaking model with pair-produced higgsinos that decay to LLPs, and each LLP subsequently decays into either a Higgs boson or a $Z$ boson.
The final state includes at least two particles that
escape direct detection, giving rise to missing transverse momentum.
No significant excess is observed above the background expectation.
The results are used to set upper limits on the cross section for higgsino pair production, up to a $\tilde\chi^0_1$ mass of 369 (704) GeV for decays with 100\% branching ratio of $\tilde\chi^0_1$ to Higgs ($Z$) bosons for a $\tilde\chi^0_1$ lifetime of 2 ns.
A model-independent limit is also set on the production of pairs of photons or electrons
with a significant delay in arrival at the calorimeter.

}

\AtlasRefCode{SUSY-2020-28}
 
\PreprintIdNumber{CERN-EP-2023-044}

\AtlasJournal{Phys.\ Rev.\ D.}
\AtlasJournalRef{\PRD 108 (2023) 012012}
\AtlasDOI{10.1103/PhysRevD.108.012012}

\hypersetup{pdftitle={ATLAS document},pdfauthor={The ATLAS Collaboration}}
 
\begin{document}
 
\maketitle

\section{Introduction}
\label{sec:intro}
 
The Standard Model (SM) of particle physics is a renormalizable quantum field theory that provides a framework for understanding fundamental particles and their interactions.
Predictions of the SM have been substantiated by experimental results over
decades, one highlight being the 2012
discovery of the
Higgs boson by the ATLAS and CMS experiments~\cite{HIGG-2012-27, CMS-HIG-12-028} at the
Large Hadron Collider (LHC) at CERN.
However, the SM does not describe gravity, does not contain a dark matter candidate, or provide a solution to the hierarchy problem, pointing to the need for new fundamental physics.
 
Supersymmetry (SUSY)~\cite{Golfand:1971iw,Neveu:1971rx,Neveu:1971iv,Ramond:1971gb,Volkov:1973ix,Wess:1973kz,Wess:1974tw} is a well-motivated theoretical extension to the SM that offers possible
answers to many of these questions.
The theory predicts the existence of SUSY partners (sparticles) for particles in the SM. 
Each sparticle has identical quantum numbers to its SM partner, differing only by half a unit of spin.
A new symmetry called R-parity acts on supersymmetric fields, and assigns a quantum number of +1 to all SM particles and -1 to sparticles.
In R-parity-conserving SUSY models~\cite{Fayet:1976et,Fayet:1977yc,Farrar:1978xj,Fayet:1979sa,Dimopoulos:1981zb}, the lightest SUSY particle~(LSP) is stable.
At colliders, sparticles would be produced in pairs, which would then decay in cascades involving other sparticles and SM particles until two final state LSPs are produced.
 
The weak eigenstates of the SUSY partners of the Higgs and gauge bosons mix to form mass eigenstates called electroweakinos that can be electrically neutral or charged fermions.
These are respectively referred to as neutralinos (\ninoone, \ninotwo, \ninothree, \ninofour) and charginos (\cinoone, \cinotwo), with the subscripts indicating increasing mass.
These mass eigenstates are model-dependent combinations of the individual electroweakino degrees of freedom.
Of the many new particles predicted in SUSY, existing exclusions from the LHC are generally weaker for these electroweak sparticles as compared to sparticles produced via strong interactions, due in part to the low electroweak process cross sections in $pp$ collisions~\cite{atlasewk,atlasewkcompressed,atlasewkdegen,cmsewk,cmsewkcompressed,cmsewkdegen}. 
 
In gauge-mediated SUSY breaking~(GMSB) models~\cite{Dine:1981gu,AlvarezGaume:1981wy,Nappi:1982,Dine:1993yw, Dine:1994vc,Dine:1995ag}, the superpartner of the graviton called the gravitino (\gravino) is the LSP for typical model parameter values.
The weak coupling of the next-to-lightest SUSY particle (NLSP) to the gravitino LSP could
generate a lifetime of the NSLP that is non-negligible on the detector length scale, leading to
displaced NLSP decay vertices~\cite{Dine:1994vc}. In GMSB models, the
lightest neutralino \ninoone is often the NLSP. If
the combination of weak eigenstates in the \ninoone \ mass eigenstate
is mostly composed of the supersymmetric Higgs eigenstate (higgsino), then the
most likely decay modes will be $\ninoone \rightarrow H/Z \ + $ \gravino.
 
\begin{figure}[h!]
\centering
\includegraphics[width=0.4\textwidth]{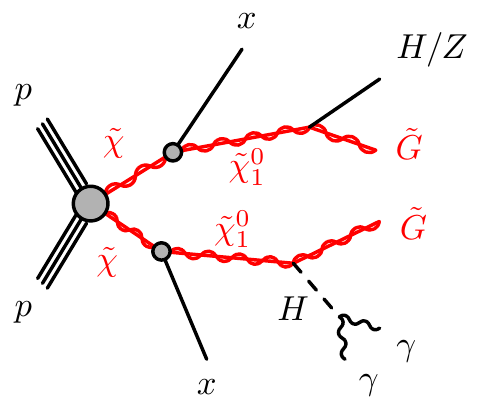}
\includegraphics[width=0.4\textwidth]{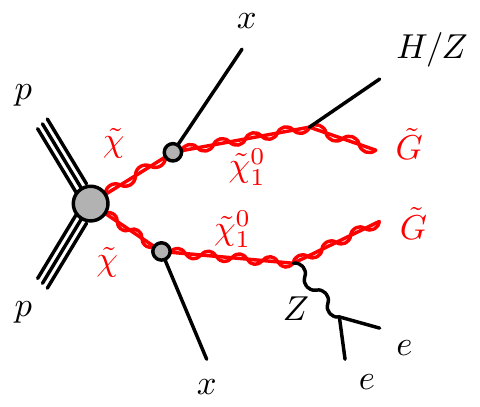}
\caption{\label{fig:feynmandiag} Feynman diagrams of the signal
process considered, targeting pair production of electroweak sparticles decaying to two light SM fermions $x$ (including all leptons and quarks except top and bottom) and a \ninoone~particle, which then decays to either a $Z$ or Higgs boson along with a \gravino. Each of the \ninoone~particles is required to decay to a Higgs (Z) boson as shown on the left (right) which decays to a diphoton (dielectron) final state. The other \ninoone is not used in the analysis, and the Higgs/$Z$ boson decays with its Standard Model branching ratio.}
\end{figure}

This search considers GMSB models with the \cinoone, \ninotwo \ and
\ninoone \ forming an almost degenerate triplet of SUSY partners of the
SM electroweak bosons,
and with the heavier charginos and neutralinos
sufficiently massive to be essentially decoupled. As shown in
the example Feynman diagrams in Figure~\ref{fig:feynmandiag},
the search focuses on direct pair production of members of the nearly degenerate triplet.
Each \ninotwo~ and \cinoone~ that is produced decays to the NLSP plus two light SM fermions (denoted by $x$), followed by the
subsequent NLSP decay via  $\ninoone \rightarrow H/Z \ + $ \gravino.
The small mass splitting between electroweakinos results in final state SM fermions with very low $p_T$, making them challenging to reconstruct and thus not useful for event selection.
The Feynman diagrams shown include both \Hyy~and \Zee~final states.
 
The analysis exploits the precision spatial and timing capabilities of the ATLAS liquid argon (LAr) electromagnetic (EM) calorimeter to achieve sensitivity to the displaced production of the SM Higgs or $Z$ boson by reconstructing the resultant
\Hyy \ or \Zee \ decays.
The main characteristic of such events is the presence of two
electromagnetic (EM) objects, either photons or electrons,
that originate
from the decay of the same LLP parent.
The EM objects are reconstructed using only EM calorimeter information, so
no attempt is made to separate between the diphoton and dielectron
final states.
These EM objects are detected with some delay compared
to prompt objects in the final state.
Given the size of the ATLAS detector and the LAr timing resolution, the requirement of LAr calorimeter measurements of the delayed EM objects restricts the sensitivity of the analysis to NLSP
lifetimes of $\mathcal{O}$(ns).
In addition, due to the opening angle between the $H/Z$ boson and
the gravitino LSP produced in the NLSP decay, the EM objects would have flight paths that are inconsistent with originating from the primary vertex (PV), and hence are called non-pointing.
Hereafter, the daughter particles will be referred to as photons, and the signature as a \textit{displaced diphoton vertex} (DDV), covering both electron and photon final states.
Precise LAr timing and spatial information is used to enhance sensitivity to this signature, and this result represents the first application of a new method to localize a displaced decay vertex position using only LAr measurements.
 
This analysis utilizes the full Run 2 ATLAS dataset of 13~TeV $pp$ collisions and is the first LHC search optimized for the DDV signature.
Previous ATLAS analyses searched for non-pointing and delayed photons produced in
long-lived NLSP decays in the datasets of $pp$ collisions collected at center-of-mass
energies of 13~TeV~\cite{ATLAS-CONF-2022-017} during Run 2 of the LHC, and at both 7~\cite{SUSY-2012-15} and 8~TeV~\cite{SUSY-2013-17} during Run 1.
No search found an excess above the SM background expectation, and results were interpreted in the context of a particular set of GMSB SUSY models. 
A recent Run~2 CMS result searching for such models
also found data agreeing with the SM prediction~\cite{CMS-EXO-19-005}.
These previous searches provide generic sensitivity to events containing one or more non-prompt photons that do not necessarily originate from a common vertex, making this result the first to exploit the correlation between e/$\gamma$ measurements expected from the GMSB delayed higgsino signal.
 
\begin{figure}[h!]
\centering
\includegraphics[width=0.7\textwidth]{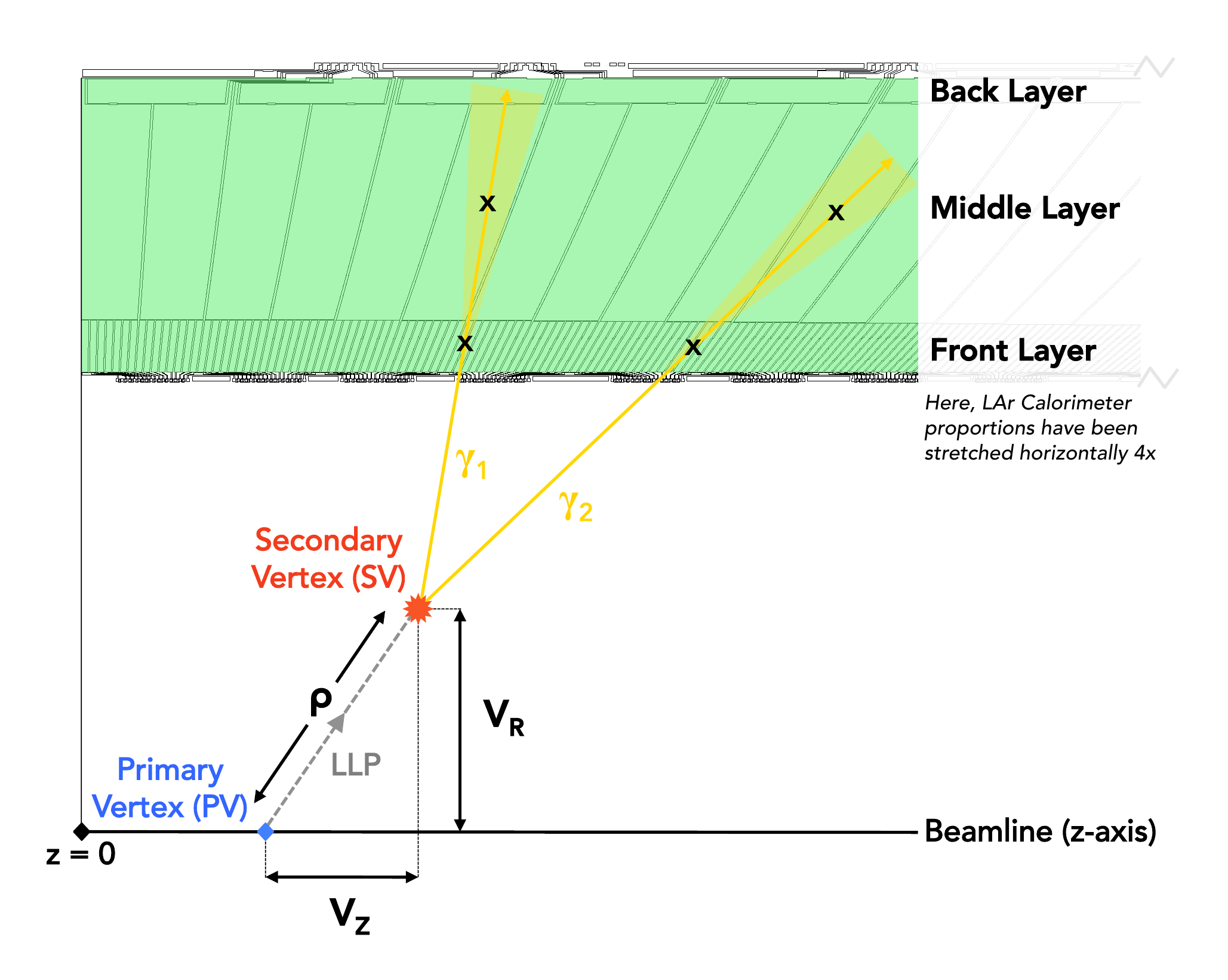}
\caption{\label{fig:cartoon_rho} Illustration of the two-dimensional calo-vertexing procedure to calculate the \vr~and \vz~discriminating variables used in the analysis, with $R$ on the y-axis and $z$ on the x-axis. The three layers of the LAr calorimeter are highlighted, along with the energy deposits left by the passage of the two daughter photons $\gamma_1$ and $\gamma_2$ of the LLP. The location of the secondary vertex (SV) is determined by the pointing measurements of the two photons. \vr~is defined as the distance in R from the SV to the beamline, and \vz~as the distance in $z$ from the SV to the PV.}
\end{figure}

The analysis considers a simplified model for signals, where the mass and lifetime of the NLSP are treated as independent parameters.
A branching ratio (BR) of unity is considered for
the combination of the two NLSP decay modes considered, namely
$\ninoone \rightarrow H/Z \ + $ \gravino, though the relative probability
of the two modes is considered a free parameter.
The $H$ and $Z$ are assumed to have their Standard Model branching ratios.
Both \Hyy \ and \Zee \ signals are reconstructed as final state photons, as electrons and photons have the same EM shower shape and thus the same EM calorimeter reconstruction.
Furthermore, both prompt and displaced track reconstruction efficiencies are low for highly displaced electrons, and the application of calorimeter-based selection can recover sensitivity to these signatures.
The photons should have a delay compared to prompt objects, and as $t = 0$ is defined as the expected value for a prompt photon from the PV of the hard collision, the photons are required to have $t > 0$.
Measurements of the trajectories of the two photons, as determined by their EM shower shapes, are used to determine a common
origin. The separation distance between this secondary vertex
candidate and the PV is calculated in the ($R$, $z$) plane, and used to
categorize the events according to the degree of displacement.
This procedure is shown diagrammatically in Figure~\ref{fig:cartoon_rho} and described in detail in Section~\ref{subsec:vertexing}.

The signal region (SR) is defined by the presence of at least two photons, as well as a high value of missing transverse momentum (\MET) resulting from the escaping gravitinos.
In addition, a few optimized kinematic selections, designed to enhance the signal-to-background ratio, are applied.
The events in the SR were kept blinded until the analysis design was finalized.
The background estimation procedure assesses the contribution of SM processes that populate the SR selection and is fully data-driven due to non-Gaussian tails in the timing and vertexing LAr measurements.
The predicted background is determined using a control region (CR) with low values of \MET, which is validated using two different orthogonal validation regions (VR).
The first VR requires intermediate values of \MET and is denoted  VR(\MET), while the second, denoted VR($t$), imposes the same \MET \ requirement as the SR but reverses the photon timing requirements by requiring \tg \ $< 0$ for each photon.
By construction the various analysis regions are orthogonal, and any signal contamination in the CR, VR(\MET), or VR($t$) was found to be negligible.
The development of the background modeling using the CR, and its validation using the VRs, were finalized before analyzing the data in the SR.
A simultaneous likelihood fit of the background model to SR data is performed using the average photon timing distribution in categories of the photon secondary vertex displacement.


\section{ATLAS detector}
\label{sec:detector}
 
\newcommand{\AtlasCoordFootnote}{
ATLAS uses a right-handed coordinate system with its origin at the nominal interaction point (IP)
in the centre of the detector and the \(z\)-axis along the beam pipe.
The \(x\)-axis points from the IP to the centre of the LHC ring,
and the \(y\)-axis points upwards.
Cylindrical coordinates \((r,\phi)\) are used in the transverse plane,
\(\phi\) being the azimuthal angle around the \(z\)-axis.
The pseudorapidity is defined in terms of the polar angle \(\theta\) as \(\eta = -\ln \tan(\theta/2)\).
Angular distance is measured in units of \(\Delta R \equiv \sqrt{(\Delta\eta)^{2} + (\Delta\phi)^{2}}\).}

The ATLAS detector~\cite{PERF-2007-01} at the LHC covers nearly the entire solid angle around the collision point.\footnote{\AtlasCoordFootnote}
It consists of an inner tracking detector surrounded by a thin superconducting solenoid, electromagnetic and hadron calorimeters,
and a muon spectrometer incorporating three large superconducting air-core toroidal magnets.
 
The inner detector system is immersed in a \SI{2}{\tesla} axial magnetic field
and provides charged-particle tracking in the range \(|\eta| < 2.5\).
The high-granularity silicon pixel detector covers the vertex region and typically provides four measurements per track,
the first hit normally being in the insertable B-layer (IBL) installed before Run~2~\cite{ATLAS-TDR-19,PIX-2018-001}.
It is followed by the silicon microstrip tracker (SCT), which provides typically eight measurements per track.
These silicon detectors are complemented by the transition radiation tracker (TRT),
which enables radially extended track reconstruction up to \(|\eta| = 2.0\).
The TRT also provides electron identification information
based on the fraction of hits (typically 30 in total) above a higher energy-deposit threshold corresponding to transition radiation.
 
The calorimeter system covers the pseudorapidity range \(|\eta| < 4.9\).
Within the region \(|\eta|< 3.2\), electromagnetic (EM) calorimetry is provided by barrel (EMB) and endcap (EMEC)
high-granularity lead/liquid-argon (LAr) calorimeters,
with an additional thin LAr presampler covering \(|\eta| < 1.8\)
to correct for energy loss in material upstream of the calorimeters.
For $\abseta < 2.5$, the LAr EM calorimeter is segmented into three layers in depth that can be used to measure the longitudinal profile of the shower.
The first layer uses highly granular ``strips'' segmented in the $\eta$ direction, with a typical transverse segmentation of $\Delta\eta \times \Delta\phi = 0.003 \times 0.1$ in the barrel, allowing for efficient discrimination between single photon showers and two overlapping showers from the decay of a $\pi^0$ meson.
The second layer has a typical transverse segmentation of $\Delta\eta \times \Delta\phi = 0.025 \times 0.025$, and collects most of the energy deposited in the calorimeter by EM showers initiated by electrons or photons.
Very high energy EM showers can leave significant energy deposits in the third layer, which can also be used to correct for energy leakage beyond the EM calorimeter.
These features of the LAr calorimeter allow it to make precise measurements of photon direction and timing, which are used to build key discriminants in the analysis as described in Section~\ref{sec:photonvars}.
Hadron calorimetry is provided by the steel/scintillator-tile calorimeter,
segmented into three barrel structures within \(|\eta| < 1.7\), and two copper/LAr hadron endcap calorimeters.
The solid angle coverage is completed with forward copper/LAr and tungsten/LAr calorimeter modules,
optimised for electromagnetic and hadronic energy measurements respectively.
 
The muon spectrometer (MS) comprises separate trigger and
high-precision tracking chambers measuring the deflection of muons in a magnetic field generated by the superconducting air-core toroidal magnets.
The field integral of the toroids ranges between \num{2.0} and \SI{6.0}{\tesla\metre}
across most of the detector.
Three layers of precision chambers, each consisting of layers of monitored drift tubes, covers the region \(|\eta| < 2.7\),
complemented by cathode-strip chambers in the forward region, where the background is highest.
The muon trigger system covers the range \(|\eta| < 2.4\) with resistive-plate chambers in the barrel, and thin-gap chambers in the endcap regions.
 
The ATLAS trigger and data acquisition system ~\cite{TRIG-2016-01} consists of a hardware-based first-level (L1) trigger followed by a software-based high-level trigger (HLT) that reduces the rate of events selected for offline storage to 1 kHz.
An extensive software suite~\cite{ATL-SOFT-PUB-2021-001} is used in the reconstruction and analysis of real
and simulated data, monitoring the detector during operation, and the trigger and data acquisition systems of the experiment.


\section{Data and Monte Carlo samples}
\label{sec:datamc}
 
This search was performed with the full Run 2 LHC dataset, collected by the ATLAS detector between 2015 and 2018.
After the application of data quality requirements~\cite{DAPR-2018-01} that ensure good working condition of all detector components, the dataset corresponds to an integrated luminosity of
$139.0 \pm 2.4$~\ifb~\cite{DAPR-2013-01,Avoni_2018}.
The recording of events used for this search was triggered by the presence of two high-\pt photons, where the full-rate trigger with the lowest available \pt~threshold is used across data-taking years~\cite{TRIG-2016-01,ATLAS:2019dpa}.
For 2015 and 2016, the trigger used requires the two photons to pass \textsc{Loose} identification (ID) selection, defined in Section~\ref{subsec:object}.
A trigger based on \textsc{Medium} ID photons became available in 2017 and 2018, and is thus used here for the data collected in those years due to its lower \pt~threshold.
Kinematic selections were imposed on the photons to ensure that the selected events lie in the fully efficient regime of the trigger, namely for the leading (subleading) photon \pt~ to be greater than 40 (30) GeV.
 
Monte Carlo event generators were used to simulate the signal targeted by this search.
Signal matrix elements were generated at leading order (LO) using
\MADGRAPH[2.7.3]
~\cite{Alwall:2014hca} with showering and hadronization 
performed by \PYTHIA[8]~\cite{Sjostrand:2007gs}.
The A14 tune~\cite{ATL-PHYS-PUB-2014-021} and the \NNPDF[2.3lo] PDF set~\cite{Ball:2014uwa} were used in the event generation.
Each event has two long-lived NLSPs and two final state $H$/\Zboson bosons.
Events were filtered such that only one $H$/\Zboson is required to decay to the desired di-$\gamma$/e resonance, and the other takes its SM branching ratios.
Therefore, four separate processes are simulated: \Hyy + $H \rightarrow$ SM, \Zee + $H \rightarrow$ SM, \Hyy + $Z \rightarrow$ SM, and \Zee + $Z \rightarrow$ SM.
The generated signals are parameterized by the mass of the NLSP, which ranges from 100~\GeV~to 725~\GeV, and its lifetime.
The mass of the gravitino LSP is set at 1 MeV, and the lightest chargino and next-to-lightest neutralino have the same mass which is set to be 1 GeV heavier than the lightest neutralino.
 
For each NLSP mass value, at least two different NLSP lifetimes, typically
2~ns and 10~ns, were simulated.
Since the distribution of particle decays follows an exponential decay curve, it is possible to reweight the shape of that curve and thus generalize to other lifetime values.
Each event was assigned a weight according to a source signal lifetime, target signal lifetime, and the decay of the event in question.
Weights for target lifetimes between 0.25 and 1000~ns were calculated using the generated signal point with the closest lifetime as the source distribution.
The signal model MC events were passed through a \GEANT~\cite{Agostinelli:2002hh}
simulation of the ATLAS detector~\cite{SOFT-2010-01} and reconstructed with the same
software~\cite{ATL-SOFT-PUB-2021-001} as used for the data.
The generation of the simulated event samples includes the effect of
multiple $pp$ interactions in the same or neighboring bunch crossings (pileup).
The effect is assessed with the inclusion of overlaid minimum-bias events, as well as the effect on the detector response due to interactions from bunch crossings before or after the one containing the hard interaction.
Events in the simulation were weighted in order to reproduce the amount of pileup
observed in the Run 2 data-taking period.

As the background estimation is fully data-driven, no simulation is required of the background processes.
Prompt SM \Zee~Monte Carlo was used to study the modeling of the specialized e/$\gamma$ variables described in Section~\ref{sec:photonvars}.
These samples were generated using \textsc{Powheg\_v1}~\cite{Frixione:2002ik} interfaced to \PYTHIA[8]~with the AZNLOCTEQ6L1 PDF/tune~\cite{ATL-PHYS-PUB-2013-017, Pumplin:2002vw}.


\section{Photon variables}
\label{sec:photonvars}
 
The capability of the ATLAS LAr calorimeter to
provide precision spatial and timing information for EM objects is essential for the reconstruction of and sensitivity to DDV events.
In addition to standard four-vector information, two key variables are used to characterize photons: the timing of the photon signal, and the pointing of its trajectory back to the beamline.
These measurements are almost completely uncorrelated for prompt backgrounds, but the signal is expected to populate higher values of both
quantities than the SM expectation, making them excellent variables to discriminate
between signal and prompt backgrounds.
Since the targeted final state has two photons that share a common
secondary vertex, the two pointing measurements are algorithmically combined into two novel vertexing variables, which describe the position of the diphoton vertex in the two-dimensional $(R, z)$ plane.
Details of the timing and vertexing calculations and their use in the analysis are provided below.
The photons and electrons used in this section are selected with the criteria enumerated in Section~\ref{subsec:object}.

\subsection{Timing}
\label{subsec:timing}
 
Photons from long-lived NLSP decays reach the LAr calorimeter with
a slight delay compared to prompt photons produced directly in the hard scattering. This
delay results
mostly from the flight time of the heavy NLSP, which would have a relativistic speed ($\beta = v/c$) that is less than 1.
In addition,
the opening angle in the NLSP decay, which causes the photon to be non-pointing,
results in the geometrical path to the calorimeter being longer than that for a prompt photon from the PV.
 
The LAr calorimeter has an accordion geometry and excellent timing resolution, with a readout which incorporates fast shaping, and a clock jitter on the readout board that is less than 20 ps.
The energy deposited by an EM object in the LAr calorimeter is measured using samples read out from the calorimeter channel at 25 ns intervals.
The arrival time is determined using the energy deposit from the second-layer calorimeter cell with the maximum energy deposit among cells in the associated EM cluster (\Ecell).
For the EM shower of an electron or photon with an energy
in the range of interest, this cell typically
contains about 20\%--50\% of the total energy deposited in the EM shower.
The energy and timing for each cell
are reconstructed by applying the
optimal filtering coefficient (OFC) algorithm~\cite{cleland} to four samples of the signal shape.
 
The time resolution $\sigma(t)$ improves as the \Ecell~increases, reaching a lower plateau at $\mathcal{O}$(10) GeV above which the resolution is flat at approximately 200 ps.
More specifically, it follows the form $\sigma(t) = p_0/\Ecell \oplus p_1$, where \Ecell \ is the cell energy, $\oplus$ denotes addition in quadrature, and parameters $p_0$ and $p_1$ are the coefficients of the so-called noise term and constant term, respectively, determined by a fit to data.
The time measurements are observed to include a correlated contribution of $\approx$ 190~ps, which agrees well with the expected spread of times from a single vertex due to their spatial distribution along the beamline.
 
An offline calibration procedure is necessary to obtain the best possible resolution on each timing measurement.
Offline corrections are determined for each interval of validity of the LAr online calibrations, of which there are 13 in the Run 2 dataset.
Timing calibration corrections are determined with a dedicated procedure that uses a large sample of \Wenu~data.
This procedure defines a time of zero as the expected measurement from a prompt photon originating at the PV, and includes corrections for offsets between channels, energy dependence, electronic crosstalk, and the position of the PV.
The calibration is validated over an independent sample of \Zee~events, which also provide a measurement of the expected resolution that is obtained by performing Gaussian fits to the time distributions in bins of cell energy.
Application of the calibration constants offline achieves a final resolution of $\mathcal{O}$(100) ps in response to high-energy e/$\gamma$ objects~\cite{ATLAS-CONF-2022-017}.

A source of early and delayed photons in data emerges through 
{satellite} bunches of protons that, due to the radio-frequency structure of the LHC accelerator and injection complex, are present in the LHC beams but separated from the main bunches by multiples of $\pm 5$~ns.
These contribute as a background process to the signal region of interest, while also allowing for an assessment of the OFC reconstruction method in data that more closely matches the expected timing distribution of the signal.
The typical population of a satellite bunch is about a factor of one thousand lower than that of the nearby nominal bunch, so collisions
between two satellite bunches are suppressed by roughly a factor
of a million. Despite their low rate, such satellite collisions
are nonetheless observable in the ATLAS data.
Figure~\ref{fig:timing_2D} shows the timing distribution of the electrons with the highest ($e_1$) and second highest ($e_2$) \pt~in data events that are subject to a selection, defined in Section~\ref{subsec:regions}, that isolates the prompt \Zee~process.
In addition to the bulk of events clustered around times of zero,
satellite collisions are seen with both electron times
around $\pm 5$~ns and also at +10~ns.
Known features of the LHC bunch structure cause a slight asymmetry between the positive and negative populations.
\begin{figure}[hbpt]
\centering
\includegraphics[width=0.48\textwidth]{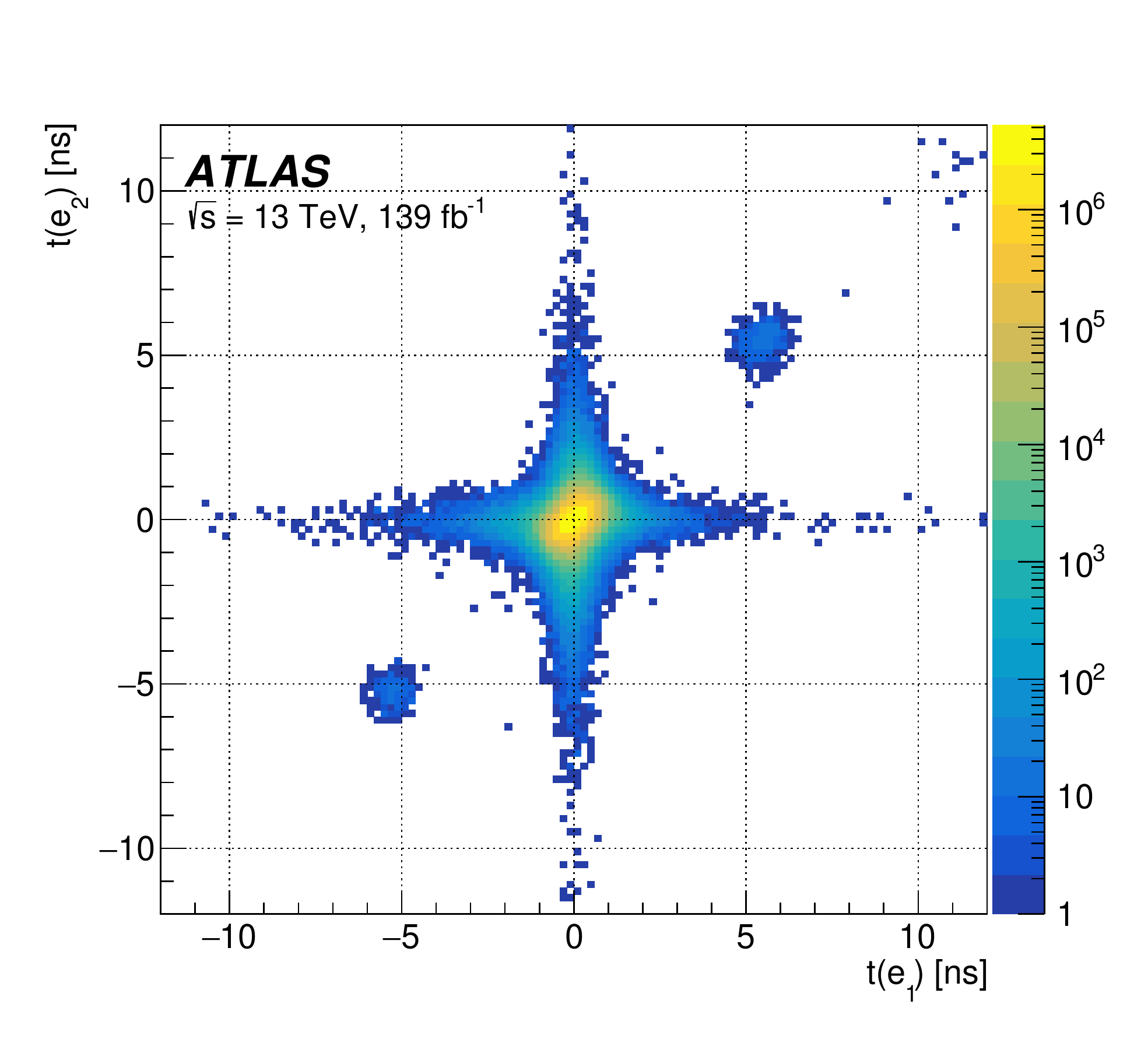}
\caption{\label{fig:timing_2D} Distribution of leading versus subleading electron LAr timing values for a \Zee~analysis selection where at least two electrons are required that have 68 GeV $<$ m$_{ee}$ $<$ 108 GeV and $|\Delta\eta(e_1,e_2)| >$ 0.1. Populations of electrons from satellite collisions are visible at $\pm 5$~ns and +10~ns.} 
\end{figure}
 
The shape of the signal timing distribution is constructed using the timing variable of the simulated signal samples.
The photon time resolution is not modeled precisely in the MC simulation, which underestimates the time resolution observed in data.
Additional resolution uncertainties are assigned to cover the observed discrepancy between radiative $Z \rightarrow \ell \ell \gamma$ events in data and those from MC simulation.
In addition to a contribution applied independently to each photon, the additional smearing includes a correlated event-level contribution to account for the impact of the spread of the actual time of the $pp$ collision, which results from the longitudinal profiles of the proton bunches along the LHC beamline.
The combined smearing contributions are tuned to match the time performance observed in data using electrons since, due to their similar EM shower developments, electrons have the same timing performance as photons.
The correlated and uncorrelated contributions to the time measurement are de-convolved by studying the times of electron-positron pairs in events from the \Zee~selection.
 
In order to exploit the correlation of the timing between two photons produced in the same parent decay, the final analysis variable is the average of the times of the two leading photons ($\gamma_1$ and $\gamma_2$), defined as \tavg~= \taveexp.
The distribution of \tavg~that is expected in the signal region as defined in Section~\ref{sec:selection}, for data and several representative simulated signals, is shown in Figure~\ref{fig:vars_t}.
The $H$ decay channels have longer timing tails due to the $m_{\gamma\gamma}$ window discussed in Section~\ref{subsec:regions}.
Because the calculation of $m_{\gamma\gamma}$ is dependent on the opening angle between the photons and thus highly displaced signals have underestimated values of the reconstructed $m_{\gamma\gamma}$, the $Z$ signals with a lower truth $m_{\gamma\gamma}$ are more likely to be cut away by the lower bound.
\begin{figure}[h!]
\centering
\includegraphics[width=0.48\textwidth]{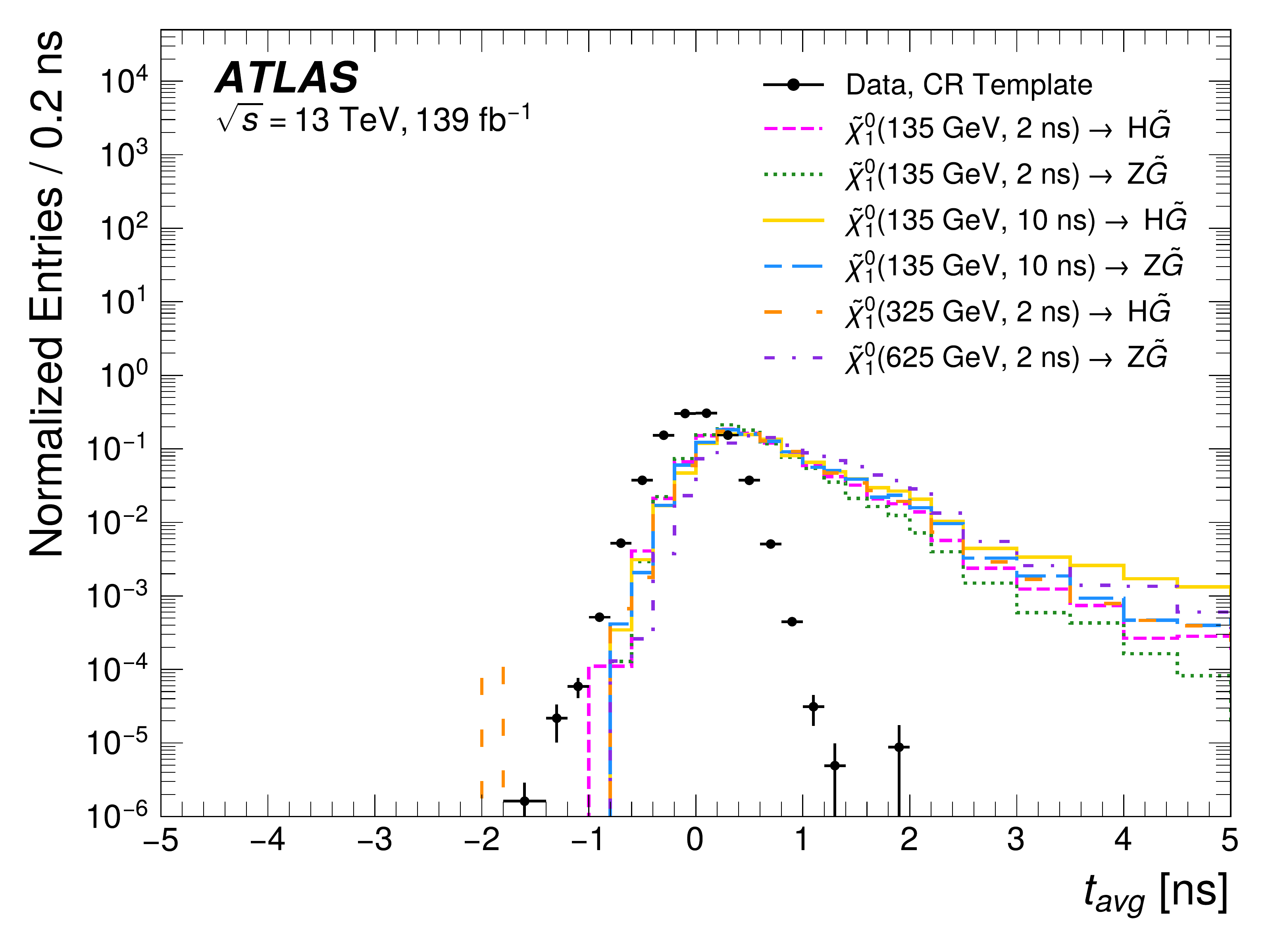}
\caption{\label{fig:vars_t} Distribution of the average timing (\tavg) for the expected background in the signal region, obtained by transforming data templates from the CR according to the background estimation procedure. Superimposed are the expected distributions for representative signal models in the SR, labeled by the \ninoone~mass (in \GeV) and lifetime (in ns), as well as the decay channel to $H$ or $Z$.}
\end{figure}
 
\subsection{Displaced Vertex Finding}
\label{subsec:vertexing}
 
The precise spatial resolution and segmentation of the LAr calorimeter provides geometrical information about the origin and direction of travel of EM objects.
In contrast to standard displaced vertex finding methods which rely on tracking information from charged particles passing through the inner detector, this analysis uses a novel method called \textit{calo-vertexing}, where the di-e/$\gamma$ production vertex is localized using only information from the LAr calorimeter.
Calo-vertexing is the only way that unconverted photons can be vertexed, while also providing enhanced acceptance for highly displaced electrons that do not have associated tracks.
 
The calo-vertexing method developed for this search uses well-established pointing variables to localize the DDV in two dimensions.
The granularity of the LAr calorimeter in $\phi$ is insufficient for unambiguous localization in this dimension, so the algorithm first rotates all photon measurements onto $\phi$=0.
It then finds each photon's pointing value, defined as the position on the beamline ($x$=$y$=0) with respect to the PV that its trajectory ``points" back to.
Pointing is determined using the signals in the first two calorimeter layers, each of which has an associated distance from the beamline $R$ and $\eta$.
A line can be drawn through the measurements from the first two calorimeter layers, ($R_1$, $\eta_1$) and ($R_2$, $\eta_2$), and extrapolated back to the beamline to determine the origin of the photon in $z$.
The resolution on the pointing measurement is $\approx$15~mm for a photon with energy of $\approx$50-100 GeV in the barrel, with good agreement between simulation and data over the observable pointing distribution~\cite{ATLAS-CONF-2022-017}.
Figure~\ref{fig:vertexingres} shows the resolution of the pointing variable \zorig as a function of \zorig, comparing \Zee~-selected data, \Zee~simulation, and signal simulation for several benchmark points.
The resolution of the predicted signal, \Zee~MC, and \Zee~ data agrees well over the range that is well-populated with events from prompt data, confirming that data and simulation are similarly modeled.
Further, the agreement of the signal and \Zee~MC across higher pointing values confirms the use of the pointing variable to describe both signal and background processes.
 
\begin{figure}
\centering
\includegraphics[width=0.48\textwidth]{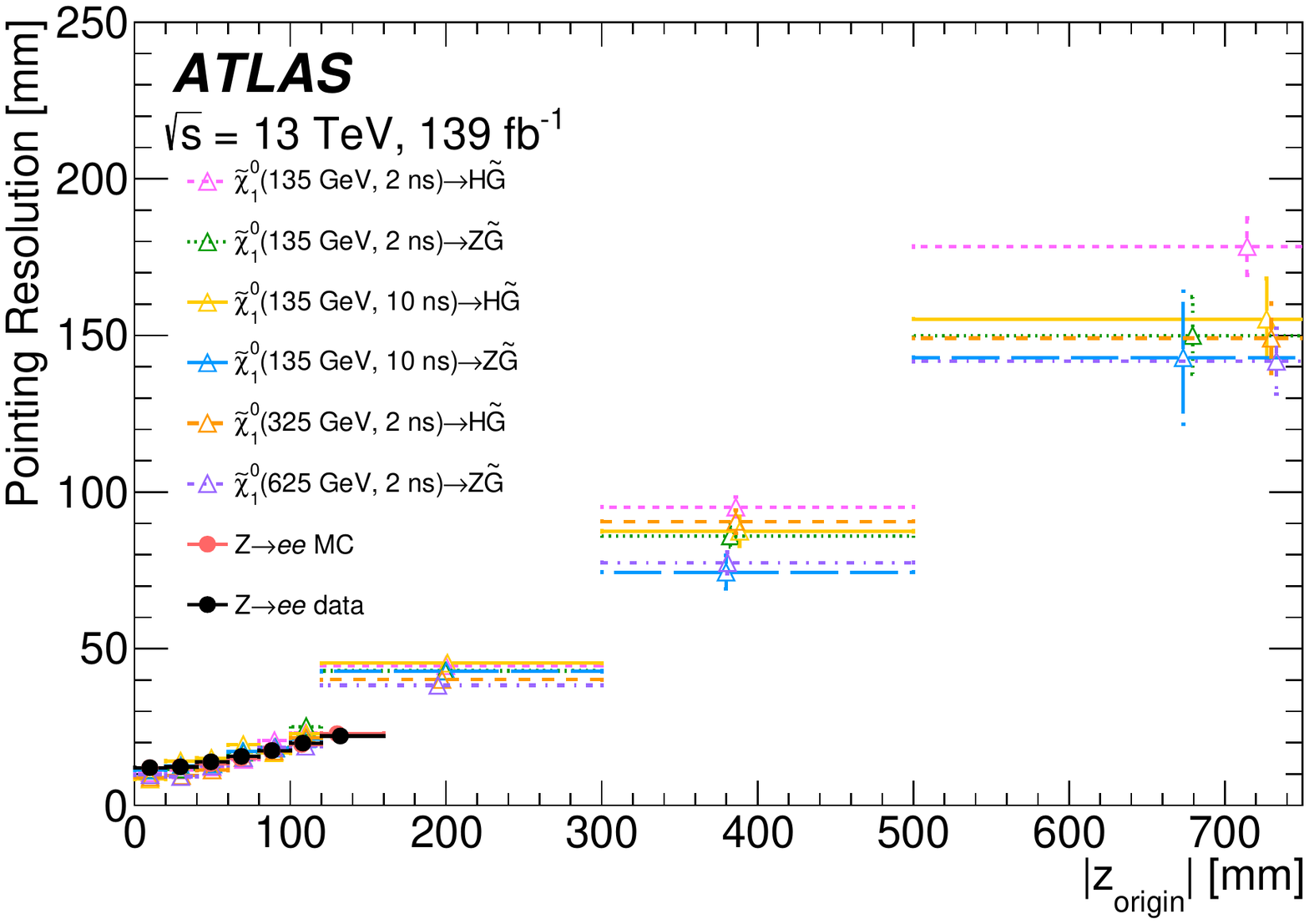}
\caption{\label{fig:vertexingres} Resolution of the photon pointing variable \zorig as a function of \zorig in units of mm. Shown are prompt \Zee~data and simulation in a selection of events with at least two electrons that have 68 GeV $<$ m$_{ee}$ $<$ 108 GeV, and $|\Delta\eta(e_1,e_2)| >$ 0.1. Overlaid for comparison are representative signals in the SR selection, labeled by the \ninoone~mass in \GeV, the \ninoone~lifetime in ns, and the decay channel to $H$ or $Z$.}
\end{figure}

The intersection of the two photon paths as determined by the pointing procedure then defines the location of the reconstructed secondary vertex candidate.
The variables \vr~and \vz~ refer to the location of the vertex along the R and $z$ axis respectively, measured with respect to the PV of the event.
Figure~\ref{fig:cartoon_rho} diagrammatically illustrates an LLP that is produced at the PV and decays after some travel distance into two photons, whose pointing measurements are used to reconstruct the SV location and thus the two vertexing variables.
 
As the pointing value will tend to be larger for photons that are produced in the decay of an LLP, and the pointing for two photons that share a common vertex will be correlated, the \vr~ and \vz~measurements are highly discriminating against prompt background.
Furthermore, the correlation between \vr~and \vz~can be exploited by summing them in quadrature, generating the variable $\rho$ = \displc \ that
is used in the analysis.
The distribution of $\rho$ that is expected in the signal region for data and several representative simulated signals is shown in Figure~\ref{fig:vars_rho}.
 
\begin{figure}[h!]
\centering
\includegraphics[width=0.48\textwidth]{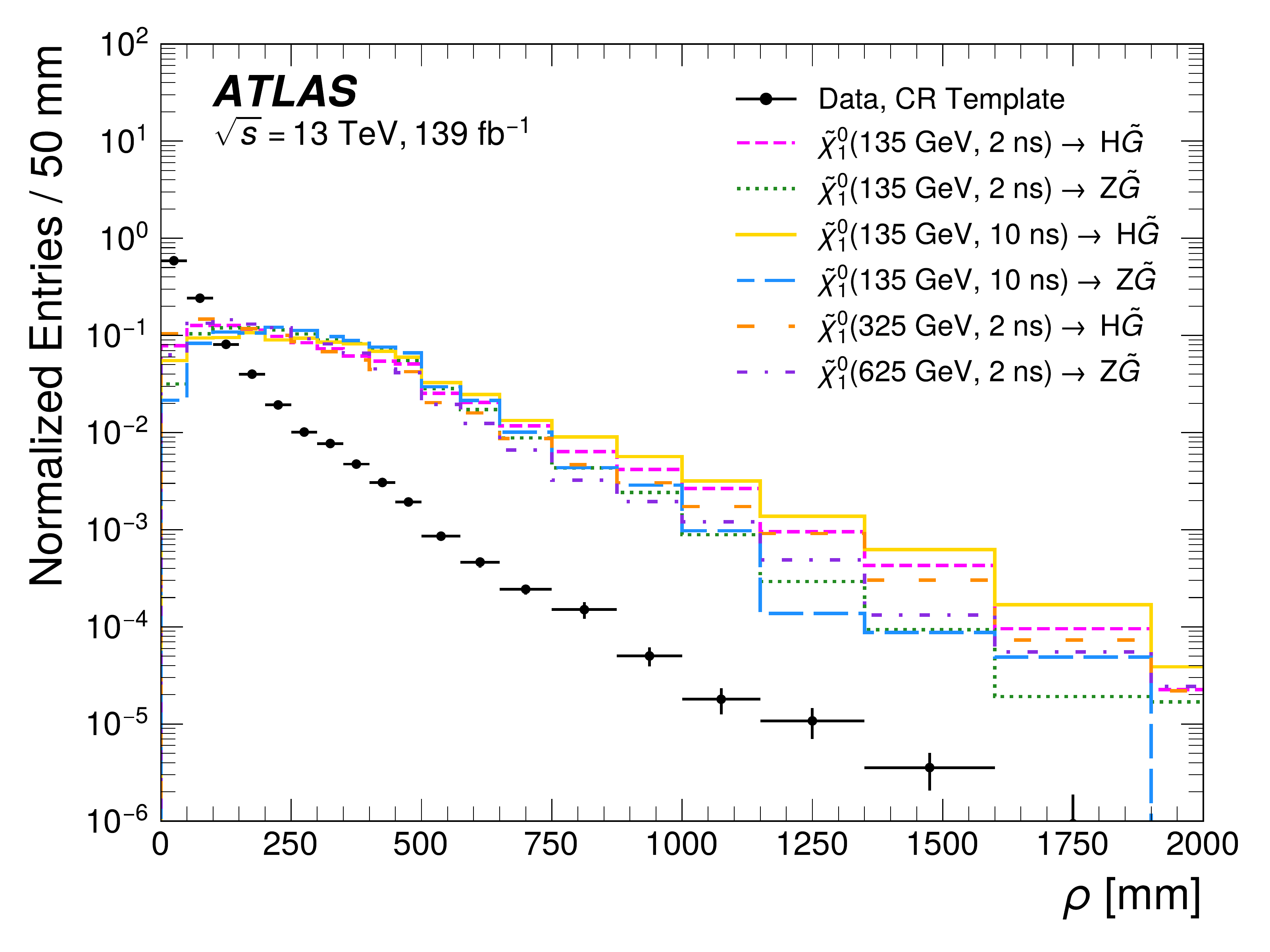}
\caption[Caption for rho]{\label{fig:vars_rho} Distribution of the displacement $\rho$ for the expected background in the signal region, obtained by transforming data templates from the CR according to the background estimation procedure. Superimposed are the expected distributions for representative signal models in the SR, labeled by the \ninoone~mass (in \GeV) and lifetime (in ns), as well as the decay channel to $H$ or $Z$. Signals with an $H$ decay have a slightly broader distribution due to sculpting of the $\rho$ distribution by the lower bound of \myy~imposed at 60 GeV\protect\footnotemark.}
\end{figure}
 
\footnotetext{The reconstruction of the invariant mass of the di-e/$\gamma$ system is biased by displacement, as the calculation of the opening angle between the two objects assumes they originate at the beamline. Decays with larger displacement leads to a greater underestimation of the \myy~value, leading to a population of events reconstructed below the truth boson mass. As the $Z$ boson mass is lower than the Higgs boson mass, more mis-reconstructed $Z$ boson invariant masses are removed by the 60 GeV lower bound.}


\section{Event selection}
\label{sec:selection}
 
Events are selected based on object quality requirements, event-level features and kinematics, as well as the timing and displacement of the two photon objects.
The selection criteria are defined based on optimization procedures that maximize signal sensitivity.

\subsection{Object Selection}
\label{subsec:object}
 
Electrons and photons are reconstructed from calorimeter signals using a dynamical, topological cell clustering-based algorithm~\cite{EGAM-2018-01}.
While the signal is reconstructed only with photons, prompt electron objects are used to define the prompt \Zee~selection.
 
Photons are required to satisfy $\pt > 10$~GeV and $|\eta|<2.37$, excluding the region $1.37 < |\eta| < 1.52$ which corresponds to the transition region between the EM barrel and endcap calorimeters.
To reduce the background of jets misreconstructed as photons, they are also subject to track- and calorimeter-based isolation requirements.
The calorimeter isolation variable is defined as the energy of calorimeter clusters around the photon candidate in the EM calorimeter in a radius of $\Delta R<0.2$. 
Additional corrections based on the leakage of photon energy outside this window, pileup, and the underlying event contribution are applied~\cite{EGAM-2018-01}.
The track-based isolation variable is defined as the scalar sum of the \pt of all tracks with $\pt>1$~GeV within $\Delta R < 0.2$ of the photon candidate.
The calorimeter (track) isolation is required to be less than 6.5\% (5\%) of the photon transverse energy.
 
A set of photon ID selections uses shower shape variables that describe the energy profiles of the EM showers in the calorimeter, and further enhance the photon efficiency while providing rejection against the background.
Three working points are employed in this analysis. Loose ID uses only variables pertaining to the second layer of the EM calorimeter and leakage into the hadronic calorimeter, which minimizes its bias against the identification of non-pointing photons.
It is applied to all baseline photon objects, including those used to calculate the \MET, and in an overlap removal procedure that prohibits the double-counting of overlapping objects in an event.
The second working point, Medium ID, shares all Loose requirements with the addition of an $\eta$-dependent selection on a first-layer shower shape variable \Eratio.
While this selection decreases signal efficiency by a few percent across the expected range of displacement values, it also offers a lower \pt~threshold by way of the Medium ID-based diphoton triggers, which was found to enhance the signal-to-background ratio overall.
Therefore, photons that are considered as potential signal objects are required to pass Medium ID.
The discrepancy between data and simulation related to the Medium identification selection is found to be negligible across displacement values, thus no systematic uncertainty on the efficiency is added.
Finally, Tight ID is used to define the photon-enriched background template described in Section~\ref{sec:background}.
It includes a variety of additional selections to ensure good rejection of fake photons.
All ID working points are defined in Ref.~\cite{EGAM-2018-01}.
 
Electron candidates must pass the same isolation and identification criteria as the photons (where the electron track is excluded from the track isolation calculation.)
The reconstructed track associated to the electron candidate must be consistent with originating at the PV, in that its longitudinal impact parameter $z_0$ and transverse impact parameter $d_0$ must satisfy $|z_0\cdot\sin\theta| < 0.5$~mm and $|d_0|/\sigma_{d_0} < 5$, respectively.
Finally, electron objects are required to have $\pt > 10$~GeV and $\aeta < 2.47$, excluding the crack region ($1.37 < \aeta < 1.52$).
 
Muons and jets only enter the analysis via overlap removal and contributions to the calculation of the missing transverse momentum \met.
Muons are reconstructed by matching tracks from the inner detector and MS subsystems.
Muons without an inner detector track in the range $2.5 < \aeta < 2.7$ but with a MS track that is compatible with the interaction point are also considered.
Muon candidates are required to have $\pt>10$~\GeV\ and $|\eta|<2.7$, and must satisfy Medium muon
identification requirements~\cite{MUON-2018-03}.
Muons are further required to satisfy calorimeter- and track-based isolation requirements~\cite{MUON-2018-03} that are 95\%--97\% efficient for muons with $\pt\in[10,60]$~\GeV\ and 99\% efficient for $\pt>60$~\GeV.
Finally, muon tracks must satisfy $|z_0\cdot\sin\theta| < 0.5$~mm and $|d_0|/\sigma_{d_0} < 3$.
 
Jets are reconstructed with \textsc{FastJet}~\cite{Fastjet} using a particle flow algorithm~\cite{PERF-2015-09} from noise-suppressed positive-energy topological clusters~\cite{PERF-2014-07,ATL-PHYS-PUB-2015-036} in the calorimeter using the anti-$k_t$ algorithm~\cite{Cacciari:2008gp} with a radius parameter R = 0.4.
Energy deposited in the calorimeter by charged particles is subtracted and replaced by the momenta of tracks that are matched to those topological clusters.
The jet four-momentum is corrected for the non-compensating calorimeter response, signal losses due to noise threshold effects, energy lost in non-instrumented regions, and contributions from the pileup. Jets are required to have $\pt > 25$~GeV and rapidity $|y| < 4.4$.
A jet-vertex-tagger (JVT) multivariate discriminant~\cite{PERF-2014-03} is applied to jets with $\pt < 60$~GeV and $\aeta < 2.4$, to suppress jets from the pileup.
 
An overlap removal procedure is performed in order to avoid double-counting objects, with photons given the highest priority.
The procedure is as follows: remove electrons overlapping with photons ($\dR < 0.4$); remove jets overlapping with photons   ($\dR < 0.4$) and those closely overlapping with electrons  ($\dR < 0.2$); remove electrons ``close to'' the remaining jets ($0.2 < \dR < 0.4$); and remove muons overlapping with photons or jets ($\dR < 0.4$).
 
The magnitude of the \met is the absolute value of the negative vector sum of the transverse momenta of the entire event and is calculated via the track-based soft term (TST) approach~\cite{PERF-2016-07}.
It uses selected photon, electron, muon, and jet objects surviving the overlap removal procedure as well as the remaining ``soft'' tracking terms that were not assigned to any of the remaining physics objects.

\subsection{Analysis Regions}
\label{subsec:regions}
 
Selected events are required to have a candidate PV for the hard scatter, reconstructed from at least two charged tracks, each with $\pT > 500$~\mev.
In case of multiple reconstructed vertices, the PV is selected as the one with the largest sum of the $\pT^2$ values of the associated tracks.
No photon pointing information is taken into consideration for PV assignment.

As the primary feature of the signal events is the presence of the DDV,
at least two photons are required in all analysis events.
Therefore, all events are subject to the offline selection of at least two photons that are matched to the triggers described in Section~\ref{sec:datamc}.
The leading and subleading photon \pt~must be greater than 40 and 30 GeV respectively, and their invariant mass \myy~must be > 60 GeV, to ensure the trigger is used in its efficiency plateau.
At least one photon must be in the barrel ($|\eta|$ < 1.37), while the other must satisfy $|\eta|$ < 1.37 or 1.52 < $|\eta|$ < 2.37.
An event cleaning procedure is applied to reject events from calorimeter noise bursts or other non-collision background, which has a negligible impact on the signal efficiency.
 
The SR is designed to select events that are consistent with the presence of a DDV.
As the DDV selections are sufficient to provide essentially background-free signal regions, the other SM boson produced by the NLSP decay is not used in the analysis in order to mitigate model-dependence of the result.
The DDV is assumed to come from the two photons that have the highest momenta in the event, whose pointing values are thus used to compute \vr~and \vz.
The two leading photons must both have $t_\gamma$ > 0, for consistency with a delayed signal, as well as $t_\gamma$ < 12 ns, to avoid contamination from adjacent bunch crossings.
They are additionally required to have an \Ecell~ measurement that is read out on \textsc{High} or \textsc{Medium} LAr gain~\cite{Buchanan:2008zza}, to ensure good performance of the offline calibration.
Further, the \Ecell~must be at least 5~GeV, a selection that balances signal acceptance with the rejection of very low-energy deposits that contaminate the timing resolution.
Selections on 0<\vr<1500 mm  and |\vz|<3740 mm ensure that the DDV is produced within the boundaries of the inner detector.
To ensure good resolution on the \vr~and \vz~reconstruction, $|\Delta \eta|$ between the two photons is required to be greater than 0.1.
As the resonance producing the DDV is assumed to be a Higgs or Z boson, \myy < 135 GeV is required.
The presence of gravitinos in the final state motivates a selection of \met > 30 GeV.
Finally, two additional selections are imposed in the SR only, on kinematic variables that were found to help isolate the signal, specifically $p_T^{\gamma \gamma} > $ 70~\GeV~and the azimuthal angle difference $\Delta \phi (\gamma_1, \gamma_2) <$ 2.4.
 
The data-driven background estimation, described in Section~\ref{sec:background}, is derived from a CR that is orthogonal to the SR by requiring \MET < 20 GeV.
Due to correlations between the photon times in the data, the timing shapes in the regions where both photons have opposite-sign timing values are slightly narrower than those in the regions with photons of same-sign times.
To ensure that the timing correlations and their impact on the analysis variable shapes match what is expected in the SR, the two photons in CR events are required to have same-sign timing values, though both ``positive-positive''
and ``negative-negative'' combinations are allowed in order to
increase the CR sample size by a factor of two.
The selection of \myy > 135 GeV in the CR is inverted with respect to the SR requirement.
Signal contamination in the control region is calculated to be less than 0.1\% for all simulated signal points.
 
Two VRs, which  are defined to be orthogonal to both the SR and CR
as well as to each other, are used to validate the background prediction.
VR(\MET) is defined in the intermediate region of the \MET~spectrum,
namely 20~\GeV < \MET < 30~GeV, providing a region of kinematic phase space that is close to that of the SR.
The inverted \myy~selection used in the CR is kept for this region.
VR($t$) uses the same \MET~ and \myy~selection as the SR, but inverts the timing selection such that only events that have two photons with negative times are allowed.
An additional prompt \Zee~-enriched region is defined to study the timing and vertexing variables, where events have at least two electrons with an invariant mass that satisfies 68 GeV $<$ m$_{ee}$ $<$ 108 GeV, and $|\Delta\eta(e_1,e_2)| >$ 0.1.
As this region is defined with electron objects and the overlap removal procedure prioritizes photons, this selection is orthogonal to all other analysis regions.
Signal contamination in the CR is less than 0.1\% across all generated signal mass and lifetime hypotheses, and less than 0.2\% in both VRs and the \Zee~selection.
 
As both signal and background have a distinct shape in \tavg and $\rho$, a shape fit is performed to exploit the entire spectrum of both variables.
The final likelihood fit is performed over the binned \tavg~distribution, for several categories of $\rho$.
Dedicated optimization studies were used to determine the binning in both variables that maximizes signal exclusion significance, while ensuring sufficient background statistics for all bins.
The projected significance of a signal hypothesis incorporating all bins was calculated for each binning considered, and optimization was performed separately for each signal point to ensure that all SR phase space is considered.
The final binning was chosen based on the discovery reach across the signal grid and
is given in Table~\ref{tab:regions}, along with a summary of all region selections discussed above.
 
\begin{table*}[hbpt]
\centering
\begin{tabular}{|c||c|c|c|c|}
\hline
Parameter & \multicolumn{4}{c|}{Preselection requirements} \\ \hline \hline
Photon multiplicity & \multicolumn{4}{c|}{> 1 }  \\ \hline
Photon $\eta$ & \multicolumn{4}{c|}{ $|\eta| < 1.37$ or 1.52 < $|\eta| < 2.37$ ($\geq$ 1 with $|\eta| < 1.37$)} \\ \hline
\ \ $\Ecell(\gamma)$ \
[\GeV] \ \ & \multicolumn{4}{c|}{$\Ecell(\gamma_1)$, \ $\Ecell(\gamma_2)$ > 5}  \\ \hline
$p_T(\gamma)$ \ [GeV] & \multicolumn{4}{c|}{$p_T(\gamma_1) > 40$, \ \ $p_T(\gamma_2) > 30$} \\ \hline
$\Delta\eta_{\gamma\gamma}$ \  & \multicolumn{4}{c|}{> 0.1 }  \\ \hline
\myy \ [GeV] & \multicolumn{4}{c|}{> 60}  \\ \hline
\vr \ [mm] & \multicolumn{4}{c|}{[0, 1500]}  \\ \hline
|\vz| \ [mm] & \multicolumn{4}{c|}{< 3740 }  \\ \hline
\tg \ [ns] & \multicolumn{4}{c|}{\tone, \ttwo \ $\in$ [-12, 12]}
\\ \hline \hline
& \multicolumn{4}{c|}{Analysis region requirements} \\
Parameter & {\ \ \ \ \ \ CR \ \ \ \ \ \ } & \ {VR(\MET)} \ & {VR($t$)} & {\ \  \ SR \ \ \ } \\ \hline \hline
\MET \ [\GeV] & \multicolumn{1}{c|}{< 20} & \multicolumn{1}{c|}{20--30} &  \multicolumn{2}{c|}{> 30}  \\  \hline
\myy[GeV] & \multicolumn{2}{c|}{> 135} & \multicolumn{2}{c|}{[60, 135]}  \\ \hline
Sign of \tg & \multicolumn{2}{c|}{\tone \ $\times$ \ \ttwo \ > \ 0 } & {\tone, \ttwo \ < \ 0} & {\tone, \ttwo \ > \ 0}
\\ \hline
$p_T^{\gamma \gamma}$ [GeV] & \multicolumn{3}{c|}{ - } & {>  70} \\ \hline
$\Delta \phi (\gamma_1, \gamma_2)$ & \multicolumn{3}{c|}{ - } & {< 2.4} \\ \hline \hline
\multicolumn{5}{|c|}{Vertexing and timing bins}  \\ \hline \hline
$\rho$ bin edges [mm] & \multicolumn{4}{c|}{[0, 80, 160, 300, 520, 2000]} \\ \hline
\tavg bin edges [ns] & \multicolumn{4}{c|}{[0, 0.2, 0.4, 0.6, 0.9, 12]} \\ \hline
\end{tabular}
\vspace*{5mm}
\caption{Optimized requirements defining the preselection, SR, CR, and both VRs. Also included are the optimized binnings for the photon vertexing ($\rho$=\displc) and average timing (\tavg) variables.}
\label{tab:regions}
\end{table*}


\section{Background estimation}
\label{sec:background}
 
SM processes do not produce a DDV with a significant invariant mass.
Background events in the SR are therefore a result of processes with either real photons that are misreconstructed to pass the DDV selection criteria (including those from satellite collisions), or other objects that fake photons.
Due to non-Gaussian tails in the LAr timing distribution, Monte Carlo generators do not give sufficiently good modeling of the data for this variable.
Therefore, a fully data-driven background estimation is used to predict the size and shape of these two background sources in the SR.
 
The low-\met CR is used to extract templates of the \tavg~shape from data.
It is known~\cite{ATLAS-CONF-2022-017,, SUSY-2013-17} that the measured pointing and timing distributions of genuine photons are narrower than those of other physics objects (specifically jets) that can be misreconstructed as photons.
To capture this shape difference, two templates are defined, one enriched in real photons and the other enriched in fake photons.
The real-enhanced photon template is defined by the CR selection given in Table~\ref{tab:regions}, in addition to a requirement that both photons satisfy the Tight identification criteria as discussed in Section~\ref{subsec:object}.
Similarly, the fake-enhanced photon template is defined as the set of events where at least one photon fails Tight ID.
 
The background in the high-\met SR is predicted by transforming and mixing these two templates in the following way.
The two templates are scaled to match the observed purity in the SR, derived from the $f_{TT}$ value obtained from data events in the SR.
The real photon CR template is multiplied by the fraction of SR events where both photons pass Tight ($f_{TT}$), and the fake photon template is weighted by $1-f_{TT}$.
The observed $f_{TT}$ in the SR ranges from approximately 0.8 in the lowest $\rho$ bin to approximately 0.5 in the highest $\rho$ bin.
This is done separately in each $\rho$ category.
Prior to purity scaling, the largest deviation in purity across the analysis regions is on the order of 10\%, and after scaling the $f_{TT}$ distributions of all regions agree within error.
Due to the dependence of calorimeter performance on the magnitude of the energy deposit, the photon timing resolution is correlated with its kinematics, specifically its energy.
A 
reweighting procedure is next applied to the templates that match each photon's \Ecell~distribution to that of the SR, which mitigates this correlation and ensures the validity of extrapolating over \MET~in the background estimation.
\Ecell-reweighting is done inclusively in timing and exclusively in $\rho$ categories, separately for each template.
The \Ecell~shape differences between the CR and SR before reweighting are on the order of 20\%, and after reweighting the distributions agree within error.
Finally, a mean correction is imposed on all data distributions to correct for small non-zero means ($\mathcal{O}$(ps)) introduced by the application of timing corrections derived from electrons to photon objects.
The correction is derived and applied separately in each \Ecell-reweighting region.
This procedure requires two features of the SR data events to be unblinded, namely the $f_{TT}$ fraction and the \Ecell~distribution.
As neither of these variables discriminates between signal and background in the region of interest and the signal contamination in the CR is below 0.1\%, this procedure doesn't affect the potential detection of signal events in the SR.
The final background template after these transformations is shown in Figures~\ref{fig:vars_t} and~\ref{fig:vars_rho} as a function of \tavg and $\rho$, respectively.


\section{Systematic uncertainties}
\label{sec:syst}
 
While the sensitivity of this search is dominated by the statistical
uncertainties in the dataset, several sources of systematic uncertainty
are considered.
These are modeled via dedicated studies and accounted for as nuisance parameters in the final statistical treatment, as described in Section~\ref{sec:stats}.
The systematic uncertainties can be either evenly distributed variations that affect only the signal normalization,
or uncertainties that impact the distribution shape of the signal and/or the background.
Systematic uncertainties are not needed on the background normalization
in each $\rho$ category, since these normalizations are implemented
as nuisance parameters that are free to float in the fit.
As expected given the limited statistical precision of the analysis region, the expected sensitivity with and without systematic uncertainties agrees to within 1\% for all mass and lifetime signal points considered.
 
\subsection{Background Shape Uncertainties}
\label{subsec:syst_bkg}
Background shape systematic uncertainties are related to the data-driven background estimation, and are assessed through variations on the estimation procedure.
 
An uncertainty exists on the purity fraction used to weight the real and fake templates to construct an SR prediction.
Since the selection criteria of an analysis region require both photons to either pass or fail \textsc{Tight} ID selection, a binomial error is assigned to the photon purity fraction.
This photon purity fraction error is then used to define maximum
deviations in the up and down directions of the photon purity fraction in bins of $\rho$.
These range from <1\% for small \tavg~and $\rho$ values, and are $<$ 10\% at their largest values in the higher bins.
A similar uncertainty exists on the \Ecell-reweighting.
The reweights are calculated from the ratio of the two-dimensional histograms between the target and the source photon \Ecell~spectra.
Therefore, the error on the reweighting is the statistical error of the ratio of the two histograms, which define the envelope of variation considered.
This effect ranges from $\mathcal{O}(1)$\% in the smallest \tavg~and $\rho$ bins to $\mathcal{O}(10)$\% for the highest.
 
Finally, a systematic uncertainty is needed to account for the extrapolation from a low-\met to a high-\met pileup profile.
This is necessary as the timing resolution for photon objects degrades with increasing pileup, due to the increased contribution of jets from pileup interactions.
To estimate this effect, the up and down variations are constructed from CR photons based on whether the average number of simultaneous interactions is less than or greater than a specified $\rho$-dependent threshold value.
These threshold values are all approximately 33, which is the average value of pileup for Run 2, and they agree within 10\% across the $\rho$ categories.
Given the limited background statistics, a smoothing procedure is applied to the pileup variation in the highest $\rho$ category by merging neighboring bins to reduce statistical fluctuations.
The size of this uncertainty is $\sim$10\% across all five $\rho$ categories.

\subsection{Signal Uncertainties}
\label{subsec:syst_sig}
 
Both normalization and shape uncertainties on the signal prediction
are considered.
These arise from experimental conditions, theoretical modeling, or analysis methodologies.
 
The instrumental systematic uncertainties which affect the signal yield are uncertainties on the photon reconstruction efficiencies, reconstruction of \MET, trigger object matching, pileup reweighting, and the integrated luminosity measurement.
The uncertainty in the integrated luminosity is $\pm $1.7\%,
evaluated using the methodology described in Ref.~\cite{DAPR-2013-01}.
The total uncertainty on the signal normalization comes from these instrumental effects and is calculated as the quadrature sum of the detector uncertainties, yielding a value of 2.5-3\% depending on the signal model.
The impact of these normalization systematic uncertainties
on the CL$_\text{s}$ value, with respect to the fit with all background shape systematic uncertainties, is less than 0.2\%.
 
Theoretical uncertainties exist on the choice of the strong coupling constant ($\alpha_{S}$),  the renormalization 
and the factorization 
scale variations, as well as the choice of PDFs.
Uncertainties are evaluated by comparing the \tavg~distribution after varying each of the parameters by a factor of two and taking an envelope of the resulting distributions.
An overall $\pm 12.6$\% normalization uncertainty is applied to all signals as a result of these theoretical modeling effects.
 
Additional uncertainties are considered that affect the distribution of
relative event yields across \tavg~bins, namely on the timing resolutions in data as compared to simulation.
The impact is measured by implementing an alternate timing smearing, adopting the same procedure of previous non-pointing photon analyses~\cite{ATLAS-CONF-2022-017}. This procedure has a negligible effect on the results.


\section{Statistical analysis}
\label{sec:stats}
 
The compatibility of the data with the background-only hypothesis is evaluated with a binned profile likelihood fit of the \tavg~ distribution of the background prediction to that observed in the data, performed simultaneously across all five $\rho$ categories.
Similarly, to evaluate the compatibility
of the data under the signal presence assumption, a fit of both signal and background templates to data is performed.
 
The likelihood function is constructed as a product of Poisson probabilities $\mathcal{P}$ for each bin of
the average timing distribution for each category.
It depends on the signal strength $\mu$, as well as the set of all nuisance parameters (NPs) with Gaussian constraints.
The $\mu$ parameter multiplies the expected signal cross-section $\sigma$ and is fully correlated across $\rho$ categories.
The normalization factor for the background in each category is modeled as a floating NP that is uncorrelated to other $\rho$ bins.
 
A profile likelihood ratio was used to perform frequentist hypothesis tests. The $p_0$-value, defined as the probability of statistical fluctuations making the background distributions appear to contain at least as much signal as the data distributions, was calculated for different signal models.
The statistical fit procedure, as well as the background estimation as derived from the CR, was validated through background-only fits to the data in the two VRs, which each have
negligible signal contamination.
Figure~\ref{fig:vrt} shows the post-fit \tavg~distribution of data and
background in the five $\rho$ categories for VR($t$), showing good agreement between data and background estimation.
 
The validity of the fit setup and background model was additionally assessed with generated pseudo-datasets. 
500 pseudo-datasets were generated by drawing events from the CR, and applying the background estimation transformations described in Section~\ref{sec:background} to match the pseudo-dataset timing shape to that of the targeted VR.
The number of events for each pseudo-dataset was fixed to the expected yield in the signal region.
This process was performed separately for both VR($t$) and VR(\MET), and a background-only fit was performed for each of the 500 pseudo-datasets to obtain a distribution of $p_0$ values for the background-only hypothesis.
The mean of the obtained $p_0$ distribution was 0.53 and 0.54 for VR($t$) and VR(\MET) respectively, reflecting good behavior of the background model in each fit.

\begin{figure}[hbpt]
\begin{centering}
\includegraphics[width=0.45\textwidth]{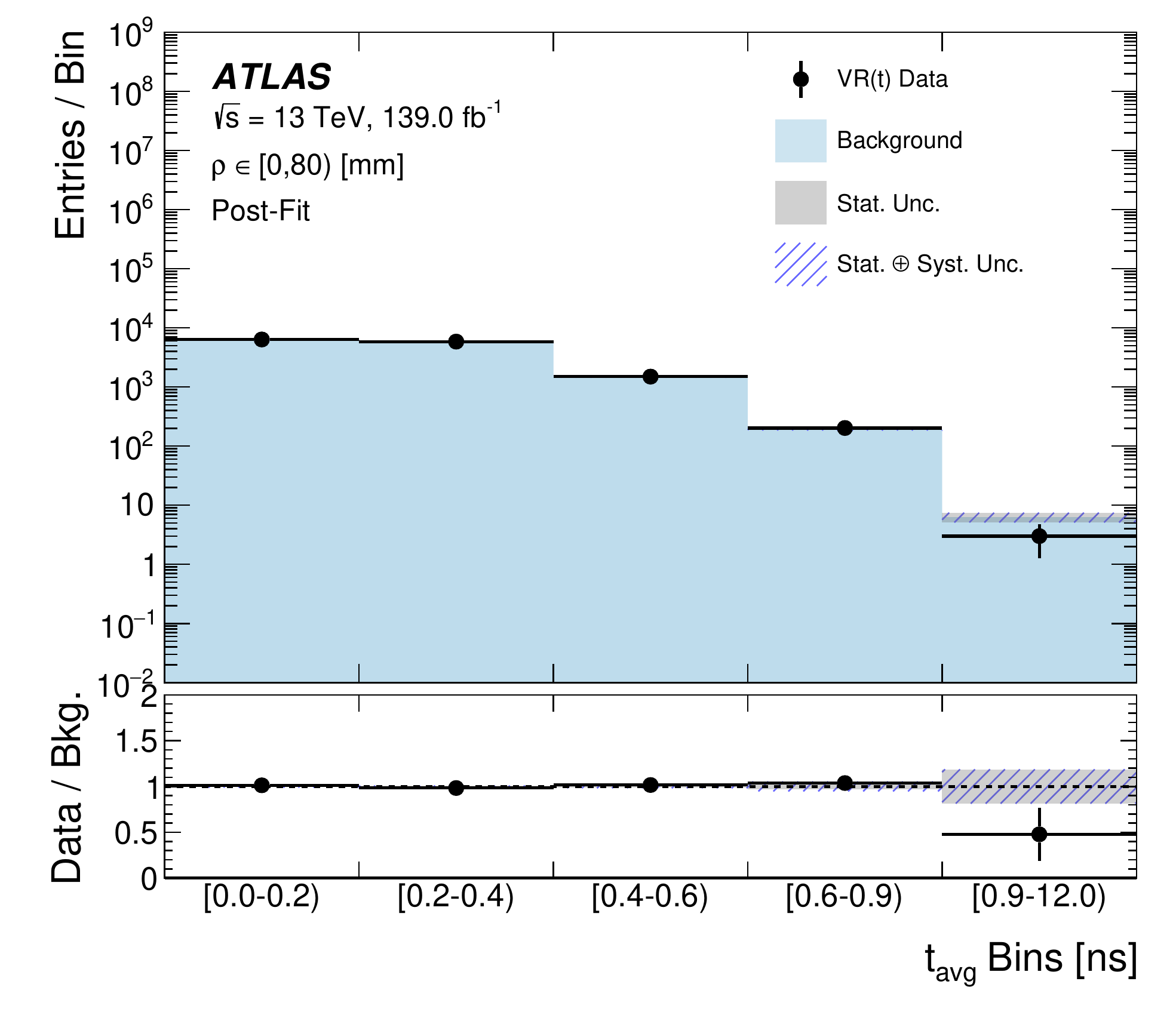}
\includegraphics[width=0.45\textwidth]{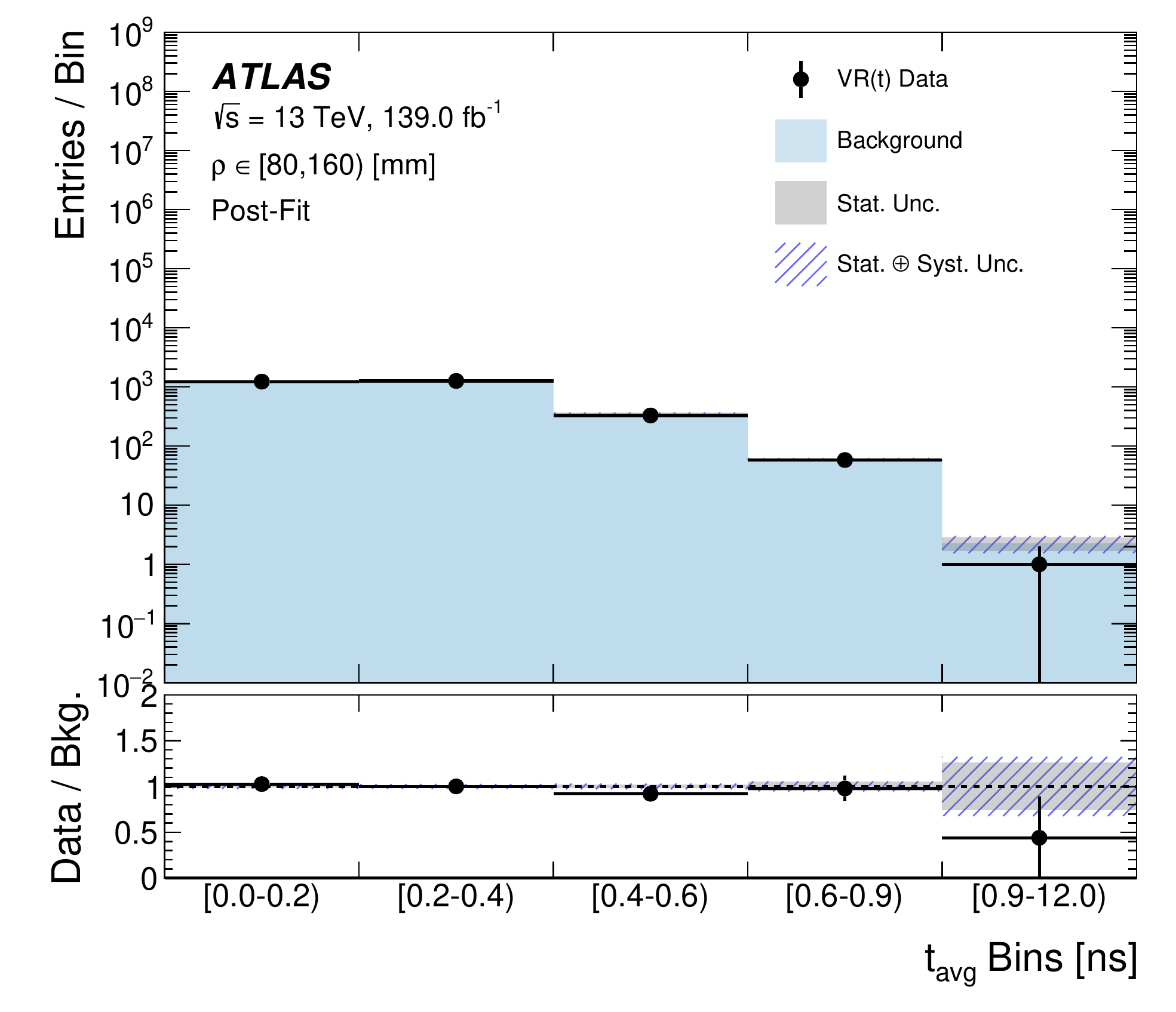}
\includegraphics[width=0.45\textwidth]{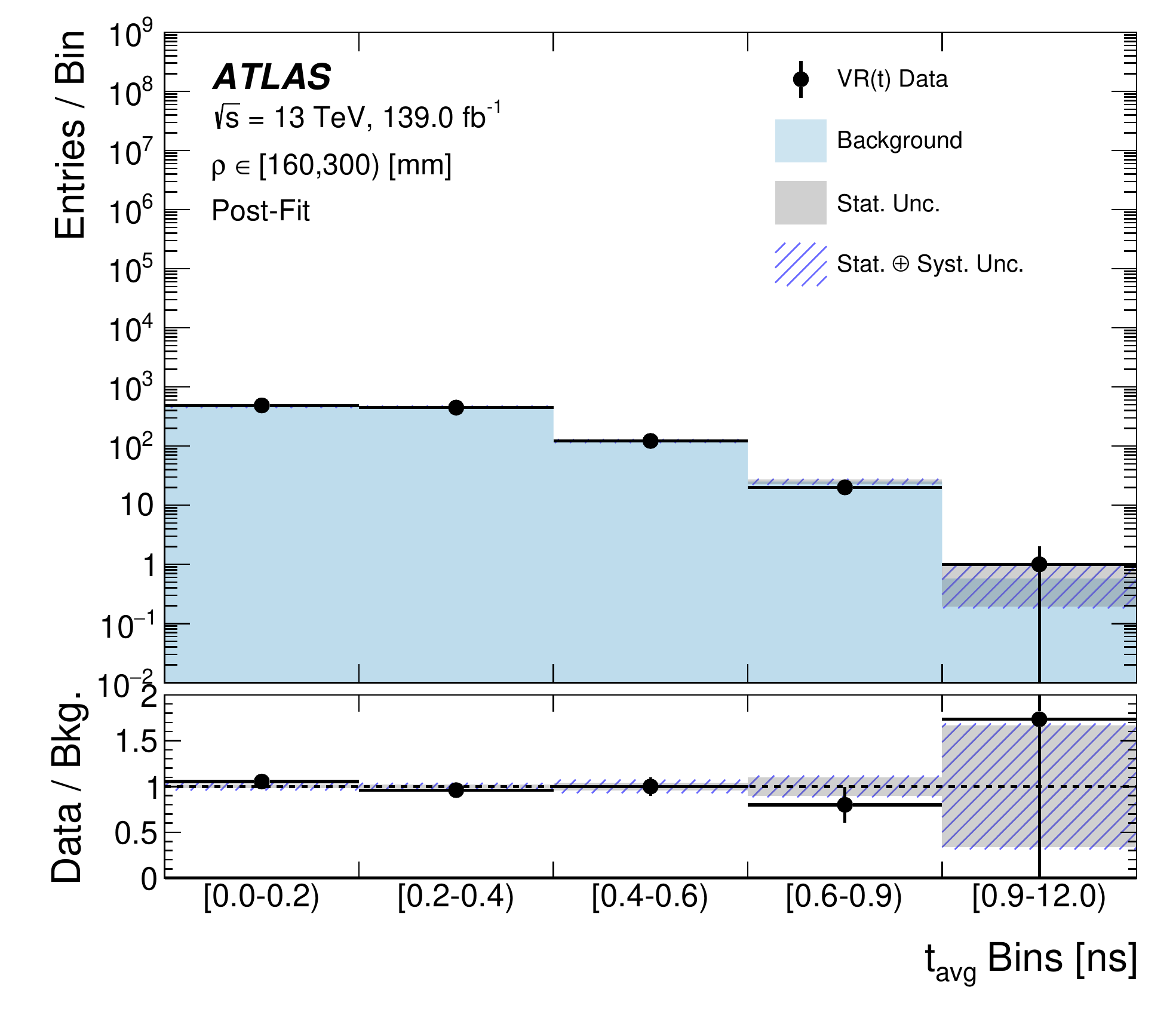}
\includegraphics[width=0.45\textwidth]{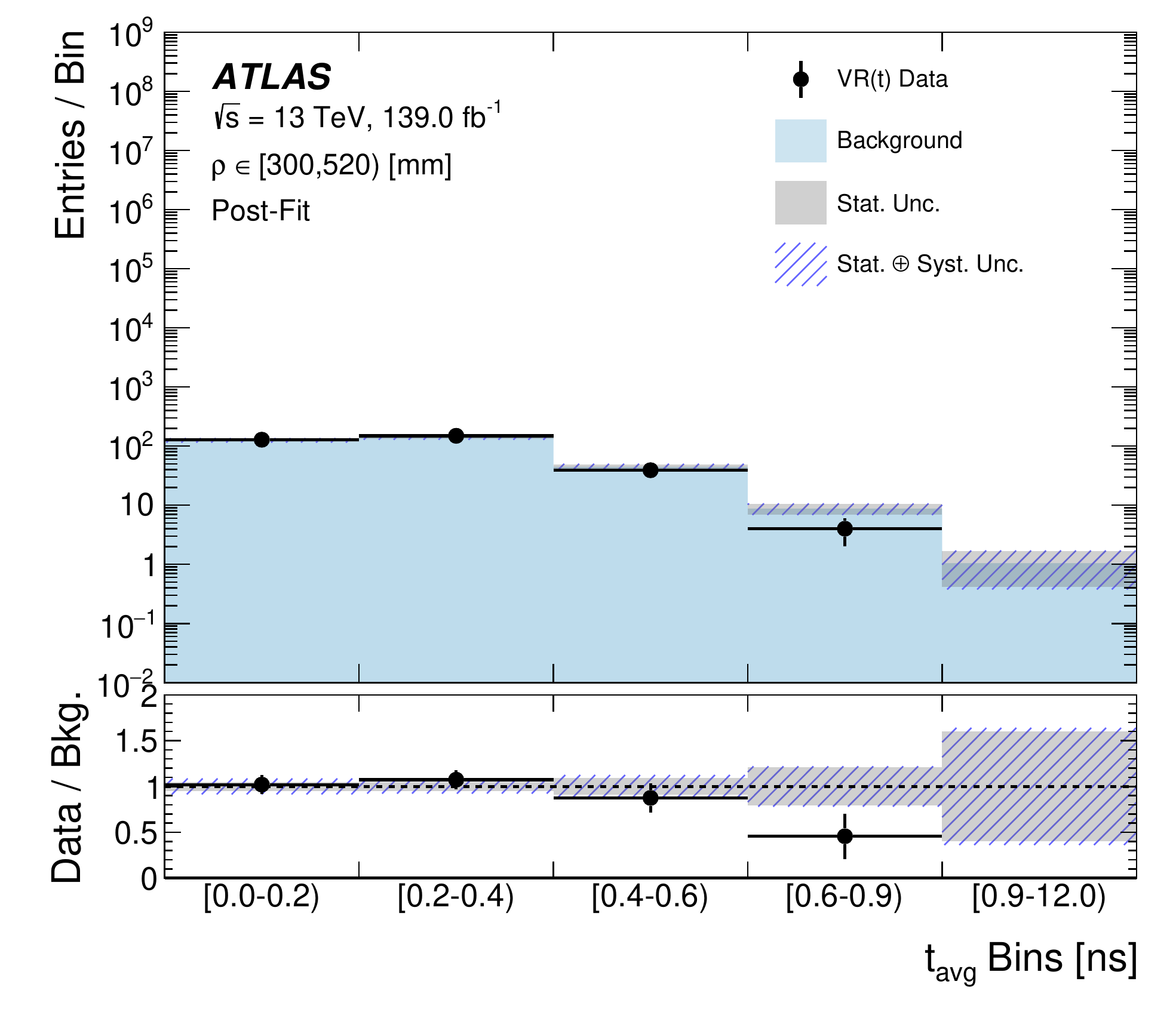}
\includegraphics[width=0.45\textwidth]{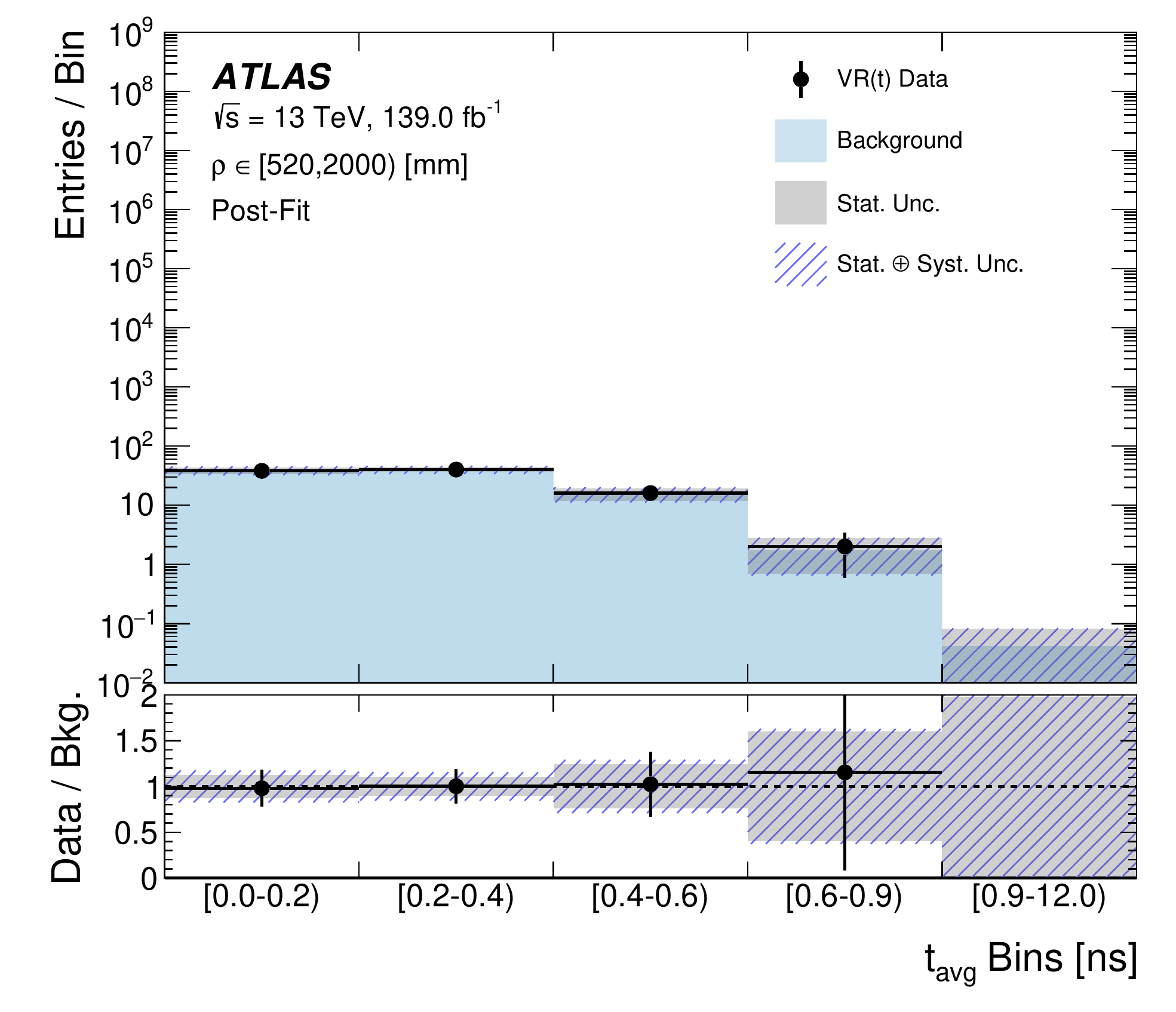}
\caption{ Average timing distributions for VR($t$) data and the estimated background as determined by the background-only fit, in each of the five exclusive $\rho$ categories.
\label{fig:vrt}}
\end{centering}
\end{figure}
 
Further validation was achieved via signal-plus-background fits for spurious signal and signal injection tests.
The spurious signal was studied via fits of the background expectation to data in the signal-depleted VRs, ensuring that the fit does not find a signal if none is present.
A conservative upper bound on the spurious signal was obtained using the signal point in the grid to which the analysis is most sensitive, which is \ninoone(135 GeV, 2 ns).
The maximum fitted significance of the aforementioned signal point across all three VRs is 0.8$\sigma$.
In the signal injection tests, the fit was performed to a pseudo-dataset consisting of CR events transformed to match the SR timing distribution shape, to which a signal template was added with varying significance.
The fitted signal $\mu$ is compared to the injected $\mu$ and they are found to be the same within statistical error across the signal grid.
The relatively similar shape of the various signal points in both \tavg~and $\rho$ leads to similar signal-plus-background results across the grid, which also serves as validation that the systematic uncertainties are appropriate for the fit.


\section{Results}
\label{sec:results}
 
Figure~\ref{fig:sr} shows the \tavg~distributions of SR data and predicted background as determined by a background-only fit with all systematic uncertainties.
The observed data are generally found to be in good agreement with the expected background.
The largest deviation occurs in the highest $\rho$ category and highest \tavg~bin, where a single event is observed with $\rho = 560$~mm, a leading photon time of 5.82~ns, and a subleading photon time of 0.45~ns.
Such an event is highly incompatible with the signal hypothesis of this search, which should have similarly delayed times for both photons.
 
The timing of the leading photon around 5~ns and the subleading photon closer to 0~ns makes this event a likely satellite collision candidate, with the leading photon coming from
a satellite collision and the second photon arising from an overlaid in-time collision.
As a known source of background, events with photons arising from satellite collisions are covered by the data-driven background estimation described in Section~\ref{sec:background}, and are thus included in the prediction of the event yield in the SR.
A simple estimate for the specific contribution of satellite collisions to the total predicted background in the SR is obtained by determining the number of satellite events present in the CR, defined by requiring photon times to be between 4.5 and 6.5 ns inclusively in $\rho$.
Since the presence of satellite collisions is uncorrelated to photon kinematics, a prediction for the SR can be calculated by scaling this number by the ratio of events in the SR to CR.
This procedure predicts that $0.5 \pm 0.3$ satellite events should be observed in the $\rho$-inclusive SR, where no systematic uncertainties are considered, thus providing further context for the likelihood of this single event observation.
 
\begin{figure}[hbpt]
\begin{centering}
\includegraphics[width=0.45\textwidth]{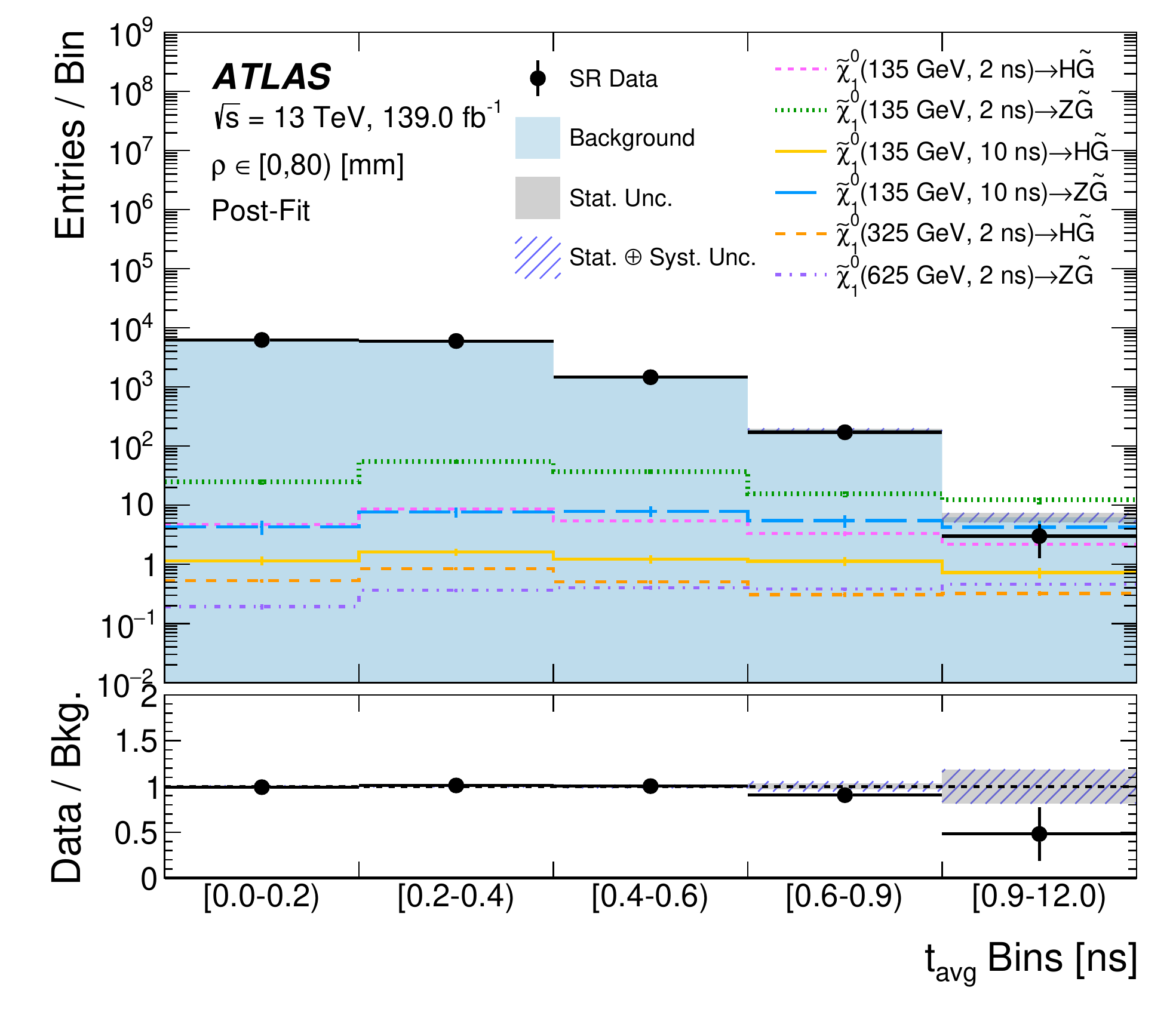}
\includegraphics[width=0.45\textwidth]{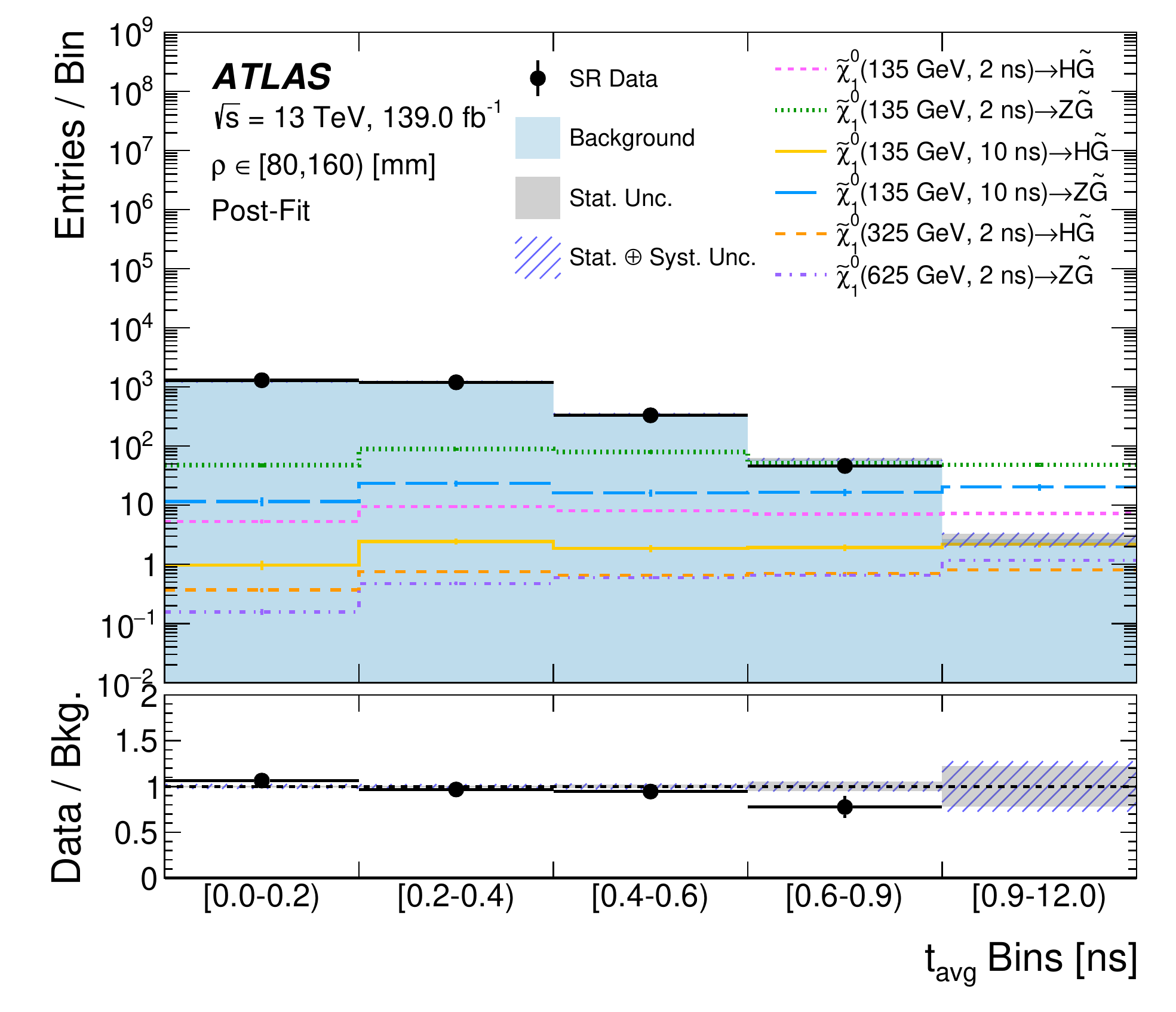}
\includegraphics[width=0.45\textwidth]{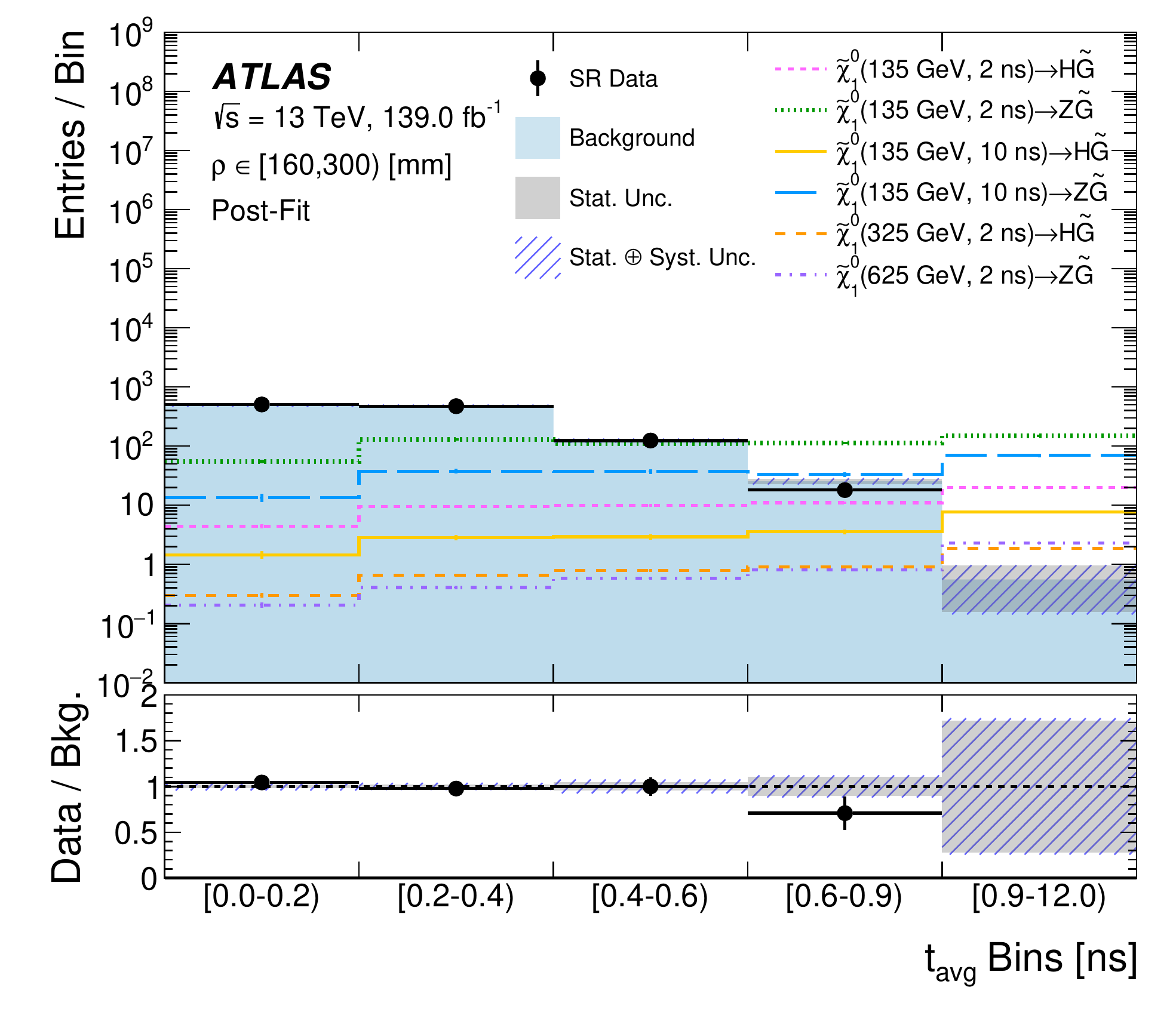}
\includegraphics[width=0.45\textwidth]{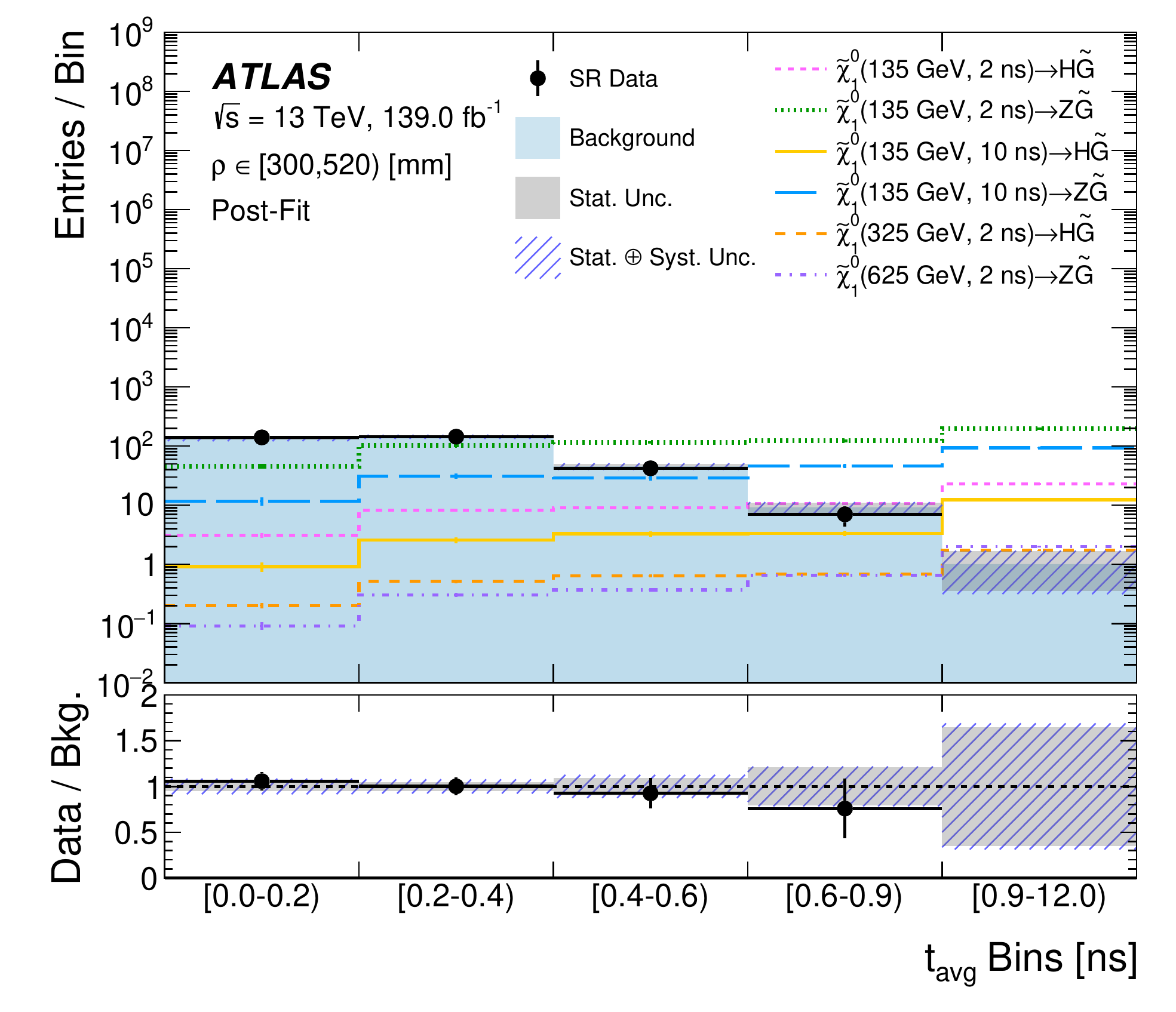}
\includegraphics[width=0.45\textwidth]{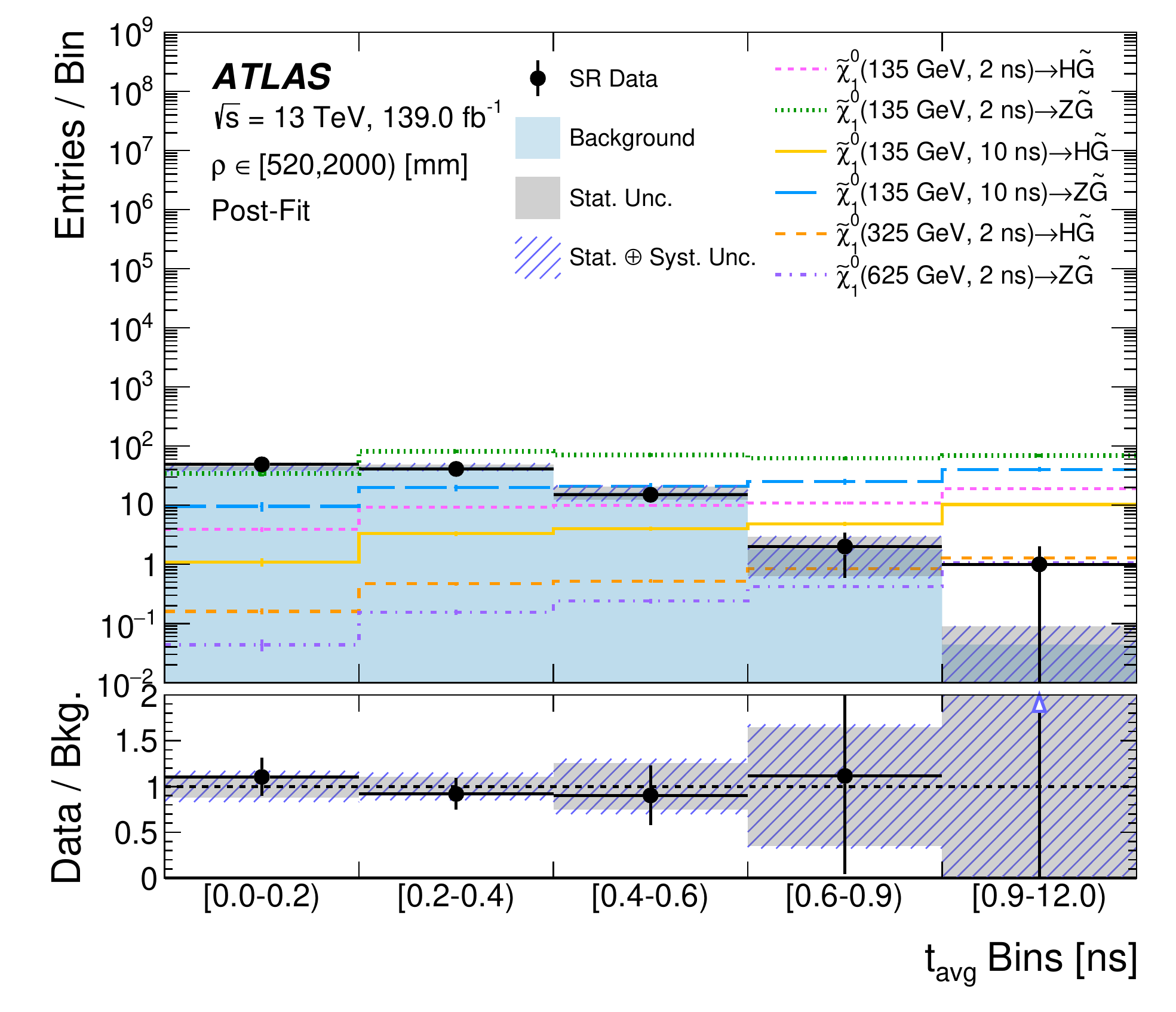}
\caption{Average timing distributions for SR data and the estimated background as determined by the background-only fit, in each of the five exclusive $\rho$ categories. For comparison, the expected timing shapes for a few different signal models are superimposed, with each model labeled by the values of the \ninoone \ mass and
lifetime, as well as decay mode.
To provide some indication of the variations
in signal yield and shape, three signal models are shown for
each of the \ninoone \
decay modes, namely \ninoone~$\rightarrow$~\HG \ and
\ninoone~$\rightarrow$~\ZG. The models shown include a
rather low \ninoone \ mass value of 135~GeV for lifetimes of
either 2~ns or 10~ns, and a higher \ninoone \ mass value which is
near the 95\% CL exclusion limit for each decay mode for a lifetime of 2~ns.
Each signal model is shown with
the signal normalization corresponding to a BR value of unity
for the decay mode in question.
\label{fig:sr}}
\end{centering}
\end{figure}
 
As no significant excess above the background prediction is observed in the SR data, signal-plus-background fits are performed to set upper limits at 95\% confidence on the signal production cross-section via the CL$_{\textrm{s}}$ technique~\cite{Junk:1999kv,Read:2002hq} under the asymptotic approximation~\cite{Cowan:2010js}.
Limits are presented as a function of the two parameters of the signal grid, the \ninoone~mass and lifetime, under the assumption of 100\% BR ($\mathcal{B}$) to either Higgs or Z bosons, as well as the BR of the \ninoone~to the SM Higgs boson for specific mass and lifetime hypotheses.
For the signal points with insufficient statistical precision where the asymptotic approximation breaks down, no limit is provided.
 
Limits on the cross-section $\sigma(pp \rightarrow \ninoone \ninoone$) in fb are shown in Figure~\ref{fig:limits}, as a function of both \ninoone~mass and lifetime, for both $H$ and $Z$ decay modes.
Figure~\ref{fig:2dlimit} shows the 95\% CL exclusion for the signal hypothesis in the two-dimensional (m$_{\ninoone}$, $\tau_{\ninoone}$) plane.
In these results, the limits for \Zee~decaying signals are more stringent than the \Hyy~decays, as the BR of the $Z$ boson to the dielectron final state is higher than that of the Higgs boson to two photons, so more events are able to pass the analysis selection criteria. \ninoone~masses up to 369 (704) GeV are excluded for the pure di-Higgs boson (di-$Z$) decay modes assuming a \ninoone~lifetime of 2 ns, which corresponds to a production cross-section of 124.5 (6.69) fb.
 
Limits as a function of the \ninoone~BR can also be computed by combining samples with both Higgs and $Z$ boson decays of the \ninoone, as well as the decay of the other \ninoone~in the event that does not produce the DDV.
In these interpretations, the sum of the \ninoone~BR to the Higgs and $Z$ bosons is assumed to be 1.
Figure~\ref{fig:limits_BR} shows an example of these limits as a function of $\mathcal{B}$(\ninoone $\rightarrow H + \tilde{G}$) for several mass hypotheses with a fixed \ninoone~lifetime of 2 ns.
 
\begin{figure}[hbpt]
\begin{centering}
\includegraphics[width=0.48\textwidth]{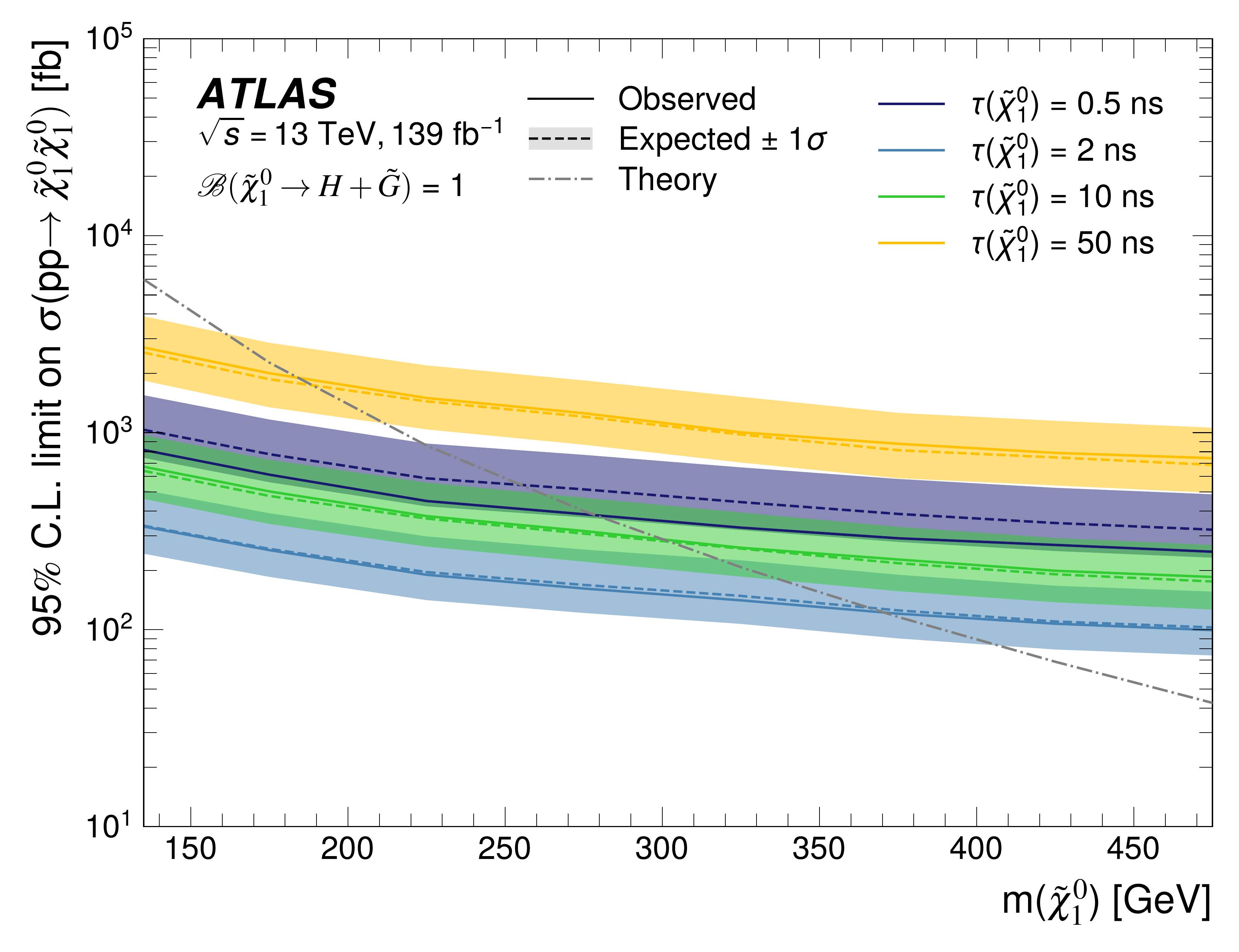}
\includegraphics[width=0.48\textwidth]{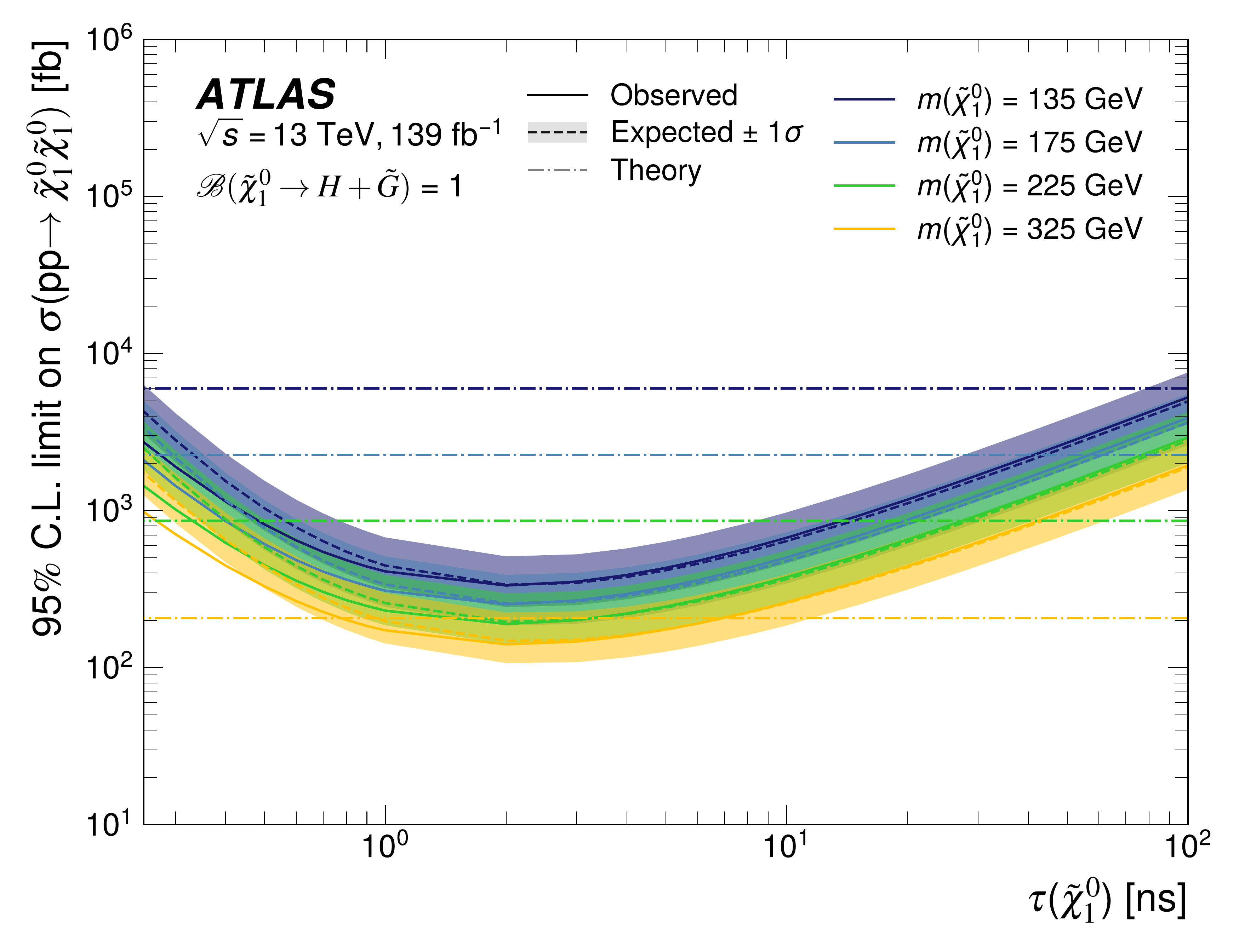}
\includegraphics[width=0.48\textwidth]{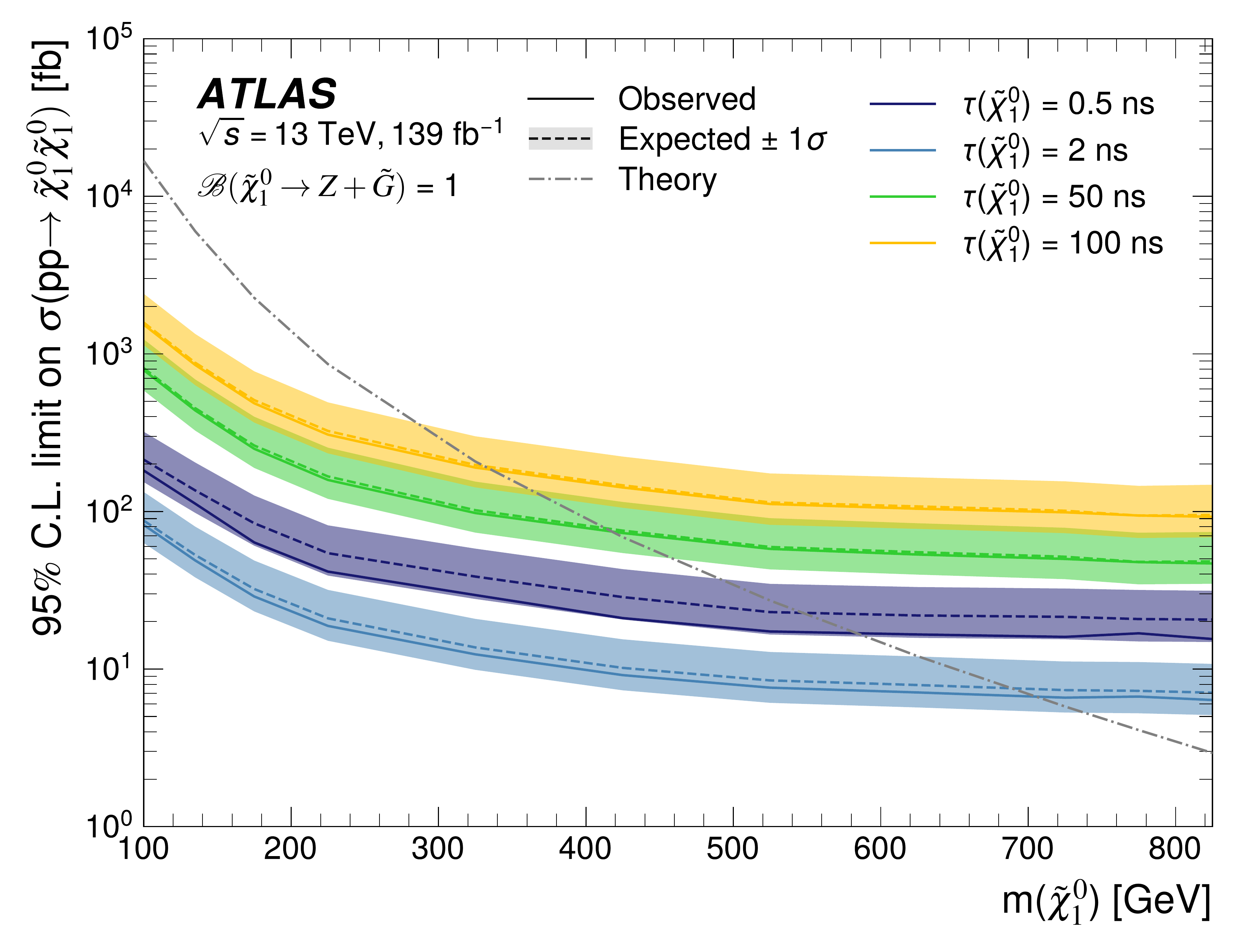}
\includegraphics[width=0.48\textwidth]{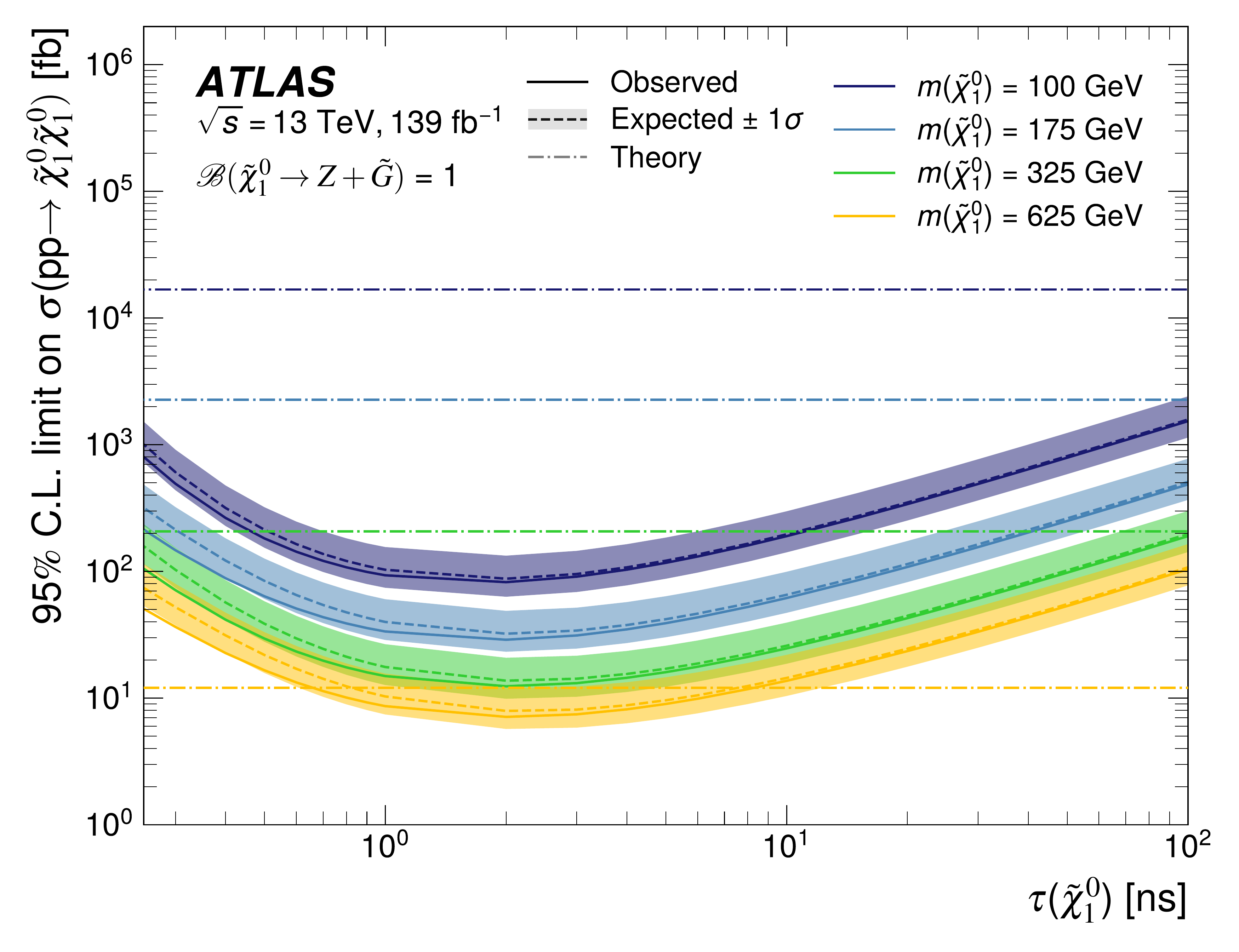}
\caption{ The 95\% CL limits on $\sigma(pp \rightarrow \ninoone \ninoone$) in fb as a function of \ninoone~mass (left) and \ninoone~lifetime (right), for the different decay modes of $\mathcal{B}$(\ninoone $\rightarrow H + \tilde{G}$) = 1 (top) and $\mathcal{B}$(\ninoone $\rightarrow Z +\tilde{G}$) = 1 (bottom). For the limits as a function of mass (lifetime), several signal models with varying lifetime (mass) are overlaid for comparison.
Included are the theoretical expectations from higgsino production for each mass hypothesis, calculated from a GMSB SUSY model that assumes nearly degenerate \ninoone, \cinoone, and \ninotwo. \label{fig:limits}}
\end{centering}
\end{figure}
 
\begin{figure}[hbpt]
\begin{centering}
\includegraphics[width=0.48\textwidth]{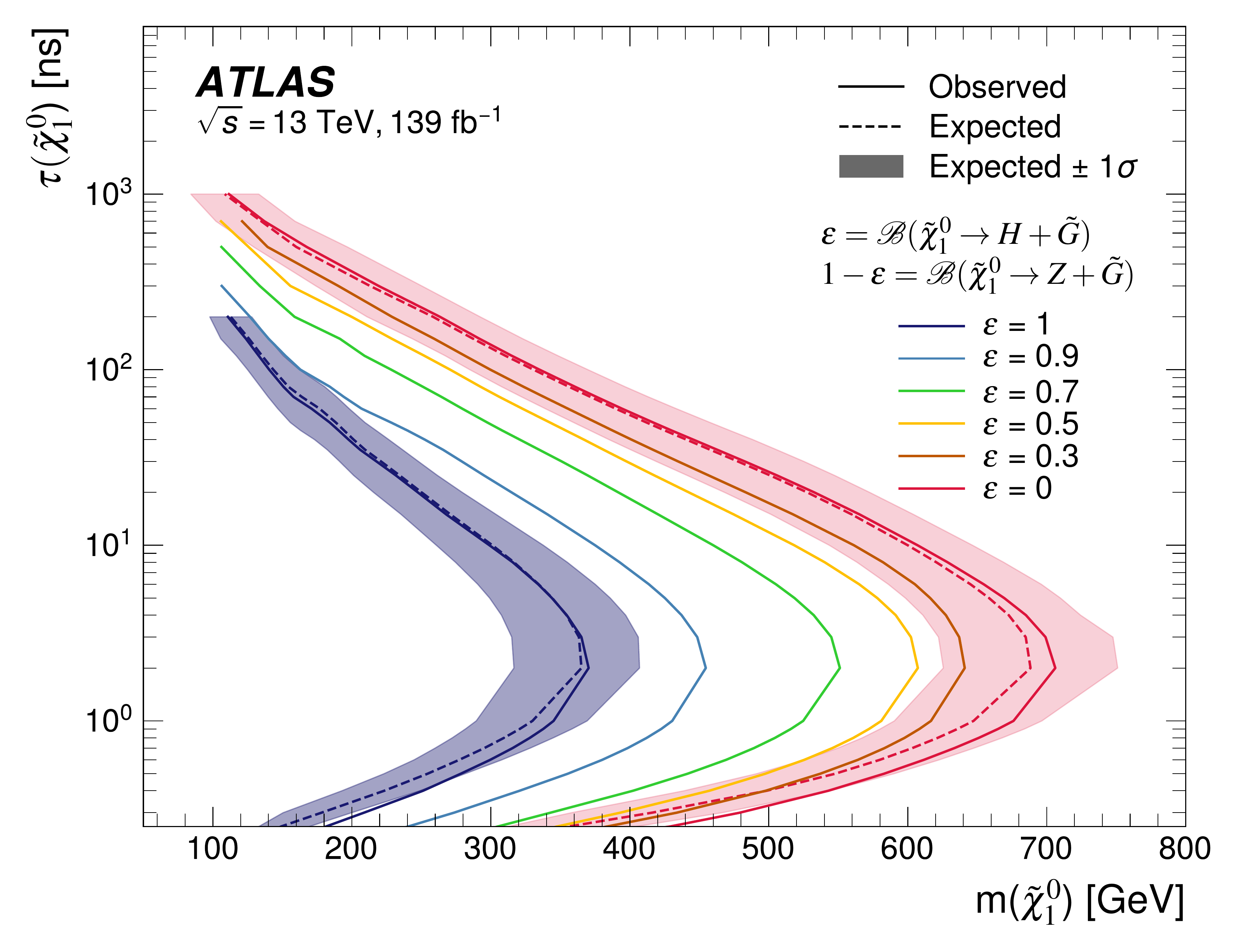}
\caption{ The 95\% CL exclusion limits on the target signal hypothesis, for \ninoone~lifetime in ns as a function of \ninoone~mass in GeV. The overlaid curves correspond to different decay hypotheses, where the assumed cross-section is for higgsino production, and the \ninoone~decays to $H + \tilde{G}$ or $Z + \tilde{G}$ such that $\mathcal{B}(H + \tilde{G}) + \mathcal{B}(Z + \tilde{G})$ = 100\%. The curve shown in red represents the decay hypothesis where the \ninoone~decays to $Z + \tilde{G}$ with 100\% branching ratio.
The curve shown in blue represents the decay hypothesis where the \ninoone~decays to $H + \tilde{G}$ with 100\% branching ratio.
\label{fig:2dlimit}}
\end{centering}
\end{figure}

\begin{figure}[hbpt]
\begin{centering}
\includegraphics[width=0.48\textwidth]{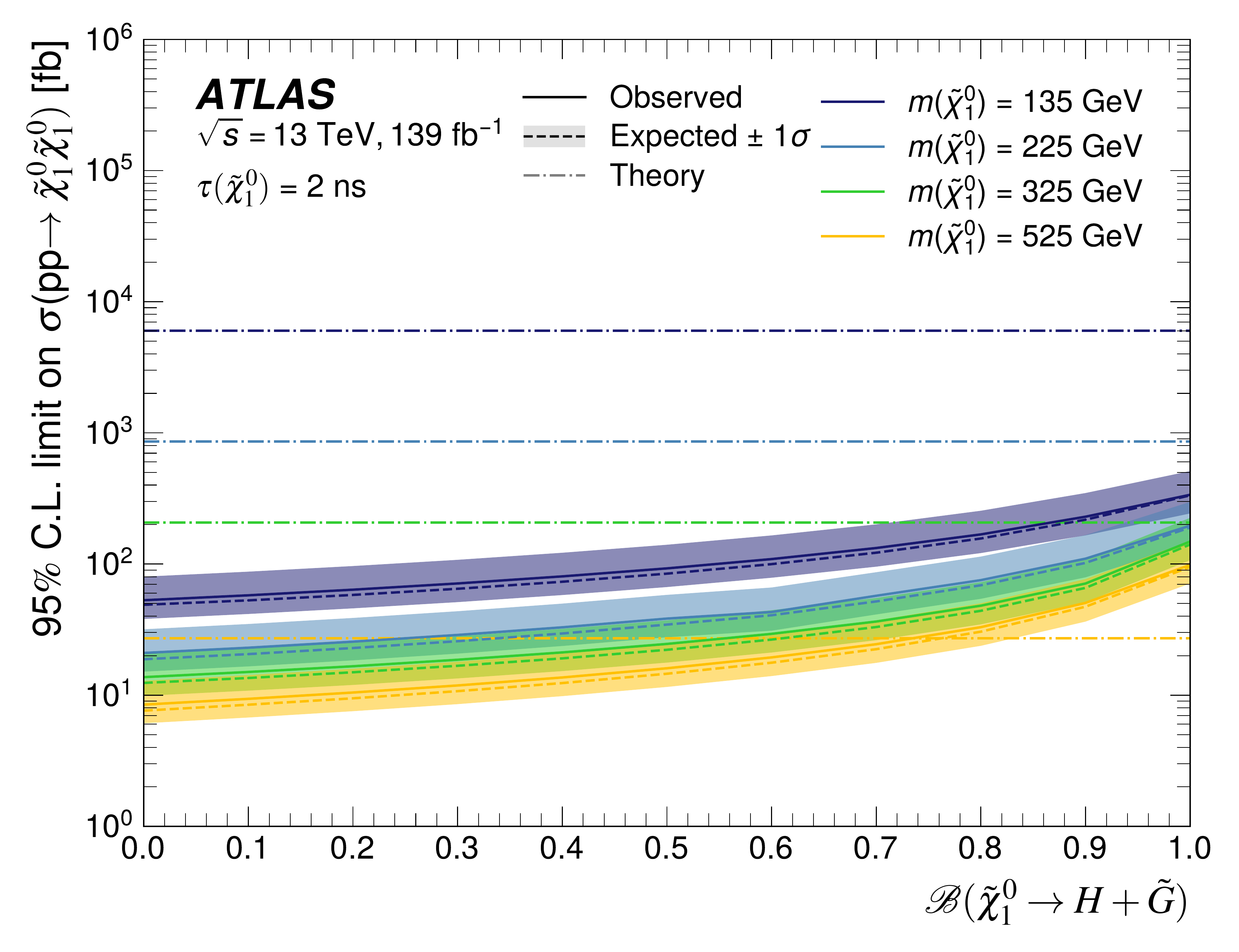}
\caption{ The 95\% CL limits on $\sigma(pp \rightarrow \ninoone \ninoone$) in fb as a function of $\ninoone$ branching ratio to the SM Higgs boson, where the assumed cross-section is for higgsino production, and $\mathcal{B}$(\ninoone $\rightarrow Z +\tilde{G}$) = 1 - $\mathcal{B}$(\ninoone $\rightarrow H + \tilde{G}$). Several signal hypotheses are overlaid that are labelled by the \ninoone~mass, all with a fixed \ninoone~lifetime of 2 ns.
Included are the theoretical expectations for each mass hypothesis, calculated from a GMSB SUSY model that assumes nearly degenerate \ninoone, \cinoone, and \ninotwo.\label{fig:limits_BR}}
\end{centering}
\end{figure}
 
In addition to the signal cross-section limits, an additional test is performed using only the final timing bin (\tavg > 0.9 ns) and no vertexing categorization.
This region enables a less model-dependent search for generic DDV signatures in data.
The results of this model-agnostic test are summarized in Table~\ref{tab:model-independent-results}.
In this bin, 10.2 $\pm$ 3.0 background events are expected and 4 are observed.
The compatibility of the observed events under the background-only hypothesis ($p_{0}$-value) and its significance (Z) are calculated.
Based on the observed and expected number of events, a 95\% CL upper limit on the number of excluded events (N$^{\textrm{excl}}$) under the background-only hypothesis is calculated.
 
{\renewcommand{\arraystretch}{1.5}
\begin{table*}[hbpt]
\centering
\begin{tabular}{|c||c|c|c|c|c|}
\hline
Selection & Data & Total Background & $p_{0}$ & Z ($\sigma$) & N$^{\textrm{excl}}$ Events \\
\hline
\hline
SR, $\tavg >$ 0.9 ns  & 4 & 10.2 $\pm$ 3.0 & 0.97 & -1.96 & 14.4 \\
\hline
\end{tabular}
\caption{Results for the model-agnostic test. The expected number of events and the uncertainty on the expected number of events are calculated using a fit to the $\rho$-inclusive \tavg distribution, with $\tavg \in$~[0.0, 0.9)~ns and extrapolated to the $\tavg \in$ [0.9, 12) ns bin. The compatibility of the observed events with the background-only expectation is given by the $p_{0}$-value, its corresponding significance Z, and the 95\% CL upper limit on the number of excluded events under the background-only hypothesis.
}
\label{tab:model-independent-results}
\end{table*}}


\section{Conclusion}
\label{sec:conclusion}
 
The first search for displaced diphoton vertices originating from the decay of a massive LLP is presented.
The data set used was recorded by the ATLAS detector at the LHC and corresponds to an integrated luminosity of 139~\ifb\ of $pp$ collisions at a center-of-mass energy of $\sqrt{s}=13$~\TeV.
Precise measurements from the ATLAS LAr calorimeter are used to select these events based on the delayed timing and displaced vertices of the final state photons.
No significant deviations are observed in the data with respect to the predicted background.
One event is observed in the highest timing bin and highest vertexing category, whose features are consistent with a source of expected background from satellite collisions.
Results are interpreted in a GMSB SUSY model as 95\% CL upper limits on the cross section of di-higgsino production scanning the \ninoone~mass, lifetime, and branching ratio to the Higgs boson, under the assumption of $\mathcal{B}(\ninoone \rightarrow H$ +\gravino) + $\mathcal{B}(\ninoone \rightarrow Z$ +\gravino)  = 1 and nearly degenerate \ninoone, \cinoone, and \ninotwo.
Limits are set across the two-dimensional space of \ninoone~mass and lifetime $\tau$, where  the \ninoone~mass is between 100 and 725 GeV and $\tau$~is between 0.25 and 1000 ns.
For a branching fraction of 100\% of \ninoone~to $H$ or $Z$ bosons, the highest excluded \ninoone~masses are 369 GeV for Higgs bosons and 704 GeV for $Z$ bosons.
Those limits are obtained for a \ninoone~lifetime of 2 ns where the analysis is most sensitive.


\section*{Acknowledgements}


We thank CERN for the very successful operation of the LHC, as well as the
support staff from our institutions without whom ATLAS could not be
operated efficiently.
 
We acknowledge the support of
ANPCyT, Argentina;
YerPhI, Armenia;
ARC, Australia;
BMWFW and FWF, Austria;
ANAS, Azerbaijan;
CNPq and FAPESP, Brazil;
NSERC, NRC and CFI, Canada;
CERN;
ANID, Chile;
CAS, MOST and NSFC, China;
Minciencias, Colombia;
MEYS CR, Czech Republic;
DNRF and DNSRC, Denmark;
IN2P3-CNRS and CEA-DRF/IRFU, France;
SRNSFG, Georgia;
BMBF, HGF and MPG, Germany;
GSRI, Greece;
RGC and Hong Kong SAR, China;
ISF and Benoziyo Center, Israel;
INFN, Italy;
MEXT and JSPS, Japan;
CNRST, Morocco;
NWO, Netherlands;
RCN, Norway;
MEiN, Poland;
FCT, Portugal;
MNE/IFA, Romania;
MESTD, Serbia;
MSSR, Slovakia;
ARRS and MIZ\v{S}, Slovenia;
DSI/NRF, South Africa;
MICINN, Spain;
SRC and Wallenberg Foundation, Sweden;
SERI, SNSF and Cantons of Bern and Geneva, Switzerland;
MOST, Taiwan;
TENMAK, T\"urkiye;
STFC, United Kingdom;
DOE and NSF, United States of America.
In addition, individual groups and members have received support from
BCKDF, CANARIE, Compute Canada and CRC, Canada;
PRIMUS 21/SCI/017 and UNCE SCI/013, Czech Republic;
COST, ERC, ERDF, Horizon 2020 and Marie Sk{\l}odowska-Curie Actions, European Union;
Investissements d'Avenir Labex, Investissements d'Avenir Idex and ANR, France;
DFG and AvH Foundation, Germany;
Herakleitos, Thales and Aristeia programmes co-financed by EU-ESF and the Greek NSRF, Greece;
BSF-NSF and MINERVA, Israel;
Norwegian Financial Mechanism 2014-2021, Norway;
NCN and NAWA, Poland;
La Caixa Banking Foundation, CERCA Programme Generalitat de Catalunya and PROMETEO and GenT Programmes Generalitat Valenciana, Spain;
G\"{o}ran Gustafssons Stiftelse, Sweden;
The Royal Society and Leverhulme Trust, United Kingdom.
 
The crucial computing support from all WLCG partners is acknowledged gratefully, in particular from CERN, the ATLAS Tier-1 facilities at TRIUMF (Canada), NDGF (Denmark, Norway, Sweden), CC-IN2P3 (France), KIT/GridKA (Germany), INFN-CNAF (Italy), NL-T1 (Netherlands), PIC (Spain), ASGC (Taiwan), RAL (UK) and BNL (USA), the Tier-2 facilities worldwide and large non-WLCG resource providers. Major contributors of computing resources are listed in Ref.~\cite{ATL-SOFT-PUB-2021-003}.


\clearpage

\printbibliography
 
\clearpage
\input{atlas_authlist}

\end{document}

%% file: atlas_authlist.tex
 
\begin{flushleft}
\hypersetup{urlcolor=black}
{\Large The ATLAS Collaboration}

\bigskip

\AtlasOrcid[0000-0002-6665-4934]{G.~Aad}$^\textrm{\scriptsize 101}$,
\AtlasOrcid[0000-0002-5888-2734]{B.~Abbott}$^\textrm{\scriptsize 119}$,
\AtlasOrcid[0000-0002-7248-3203]{D.C.~Abbott}$^\textrm{\scriptsize 102}$,
\AtlasOrcid[0000-0002-1002-1652]{K.~Abeling}$^\textrm{\scriptsize 55}$,
\AtlasOrcid[0000-0002-8496-9294]{S.H.~Abidi}$^\textrm{\scriptsize 29}$,
\AtlasOrcid[0000-0002-9987-2292]{A.~Aboulhorma}$^\textrm{\scriptsize 35e}$,
\AtlasOrcid[0000-0001-5329-6640]{H.~Abramowicz}$^\textrm{\scriptsize 150}$,
\AtlasOrcid[0000-0002-1599-2896]{H.~Abreu}$^\textrm{\scriptsize 149}$,
\AtlasOrcid[0000-0003-0403-3697]{Y.~Abulaiti}$^\textrm{\scriptsize 116}$,
\AtlasOrcid[0000-0003-0762-7204]{A.C.~Abusleme~Hoffman}$^\textrm{\scriptsize 136a}$
% ...
\bigskip
\\

% ...
$^{29}$Physics Department, Brookhaven National Laboratory, Upton NY; United States of America.\\
% ...
$^{35}$$^{(a)}$Facult\'e des Sciences Ain Chock, R\'eseau Universitaire de Physique des Hautes Energies - Universit\'e Hassan II, Casablanca;$^{(b)}$Facult\'{e} des Sciences, Universit\'{e} Ibn-Tofail, K\'{e}nitra;$^{(c)}$Facult\'e des Sciences Semlalia, Universit\'e Cadi Ayyad, LPHEA-Marrakech;$^{(d)}$LPMR, Facult\'e des Sciences, Universit\'e Mohamed Premier, Oujda;$^{(e)}$Facult\'e des sciences, Universit\'e Mohammed V, Rabat;$^{(f)}$Institute of Applied Physics, Mohammed VI Polytechnic University, Ben Guerir; Morocco.\\
% ...
$^{55}$II. Physikalisches Institut, Georg-August-Universit\"{a}t G\"ottingen, G\"ottingen; Germany.\\
% ...
$^{101}$CPPM, Aix-Marseille Universit\'e, CNRS/IN2P3, Marseille; France.\\
$^{102}$Department of Physics, University of Massachusetts, Amherst MA; United States of America.\\
% ...
$^{116}$Department of Physics, New York University, New York NY; United States of America.\\
% ...
$^{119}$Homer L. Dodge Department of Physics and Astronomy, University of Oklahoma, Norman OK; United States of America.\\
% ...
$^{136}$$^{(a)}$Departamento de F\'isica, Pontificia Universidad Cat\'olica de Chile, Santiago;$^{(b)}$Millennium Institute for Subatomic physics at high energy frontier (SAPHIR), Santiago;$^{(c)}$Instituto de Investigaci\'on Multidisciplinario en Ciencia y Tecnolog\'ia, y Departamento de F\'isica, Universidad de La Serena;$^{(d)}$Universidad Andres Bello, Department of Physics, Santiago;$^{(e)}$Instituto de Alta Investigaci\'on, Universidad de Tarapac\'a, Arica;$^{(f)}$Departamento de F\'isica, Universidad T\'ecnica Federico Santa Mar\'ia, Valpara\'iso; Chile.\\
% ...
$^{149}$Department of Physics, Technion, Israel Institute of Technology, Haifa; Israel.\\
$^{150}$Raymond and Beverly Sackler School of Physics and Astronomy, Tel Aviv University, Tel Aviv; Israel.\\
% ...
$^{a}$ Also Affiliated with an institute covered by a cooperation agreement with CERN.\\
$^{b}$ Also at An-Najah National University, Nablus; Palestine.\\
$^{c}$ Also at Borough of Manhattan Community College, City University of New York, New York NY; United States of America.\\
$^{d}$ Also at Bruno Kessler Foundation, Trento; Italy.\\
$^{e}$ Also at Center for High Energy Physics, Peking University; China.\\
$^{f}$ Also at Center for Interdisciplinary Research and Innovation (CIRI-AUTH), Thessaloniki ; Greece.\\
% ...

\end{flushleft}
